\documentclass[showpacs,twocolumn,aps,prx,longbibliography,superscriptaddress]{revtex4-2}

\usepackage{kotex}

\usepackage{graphicx}
\usepackage{physics}
\usepackage{wrapfig}
\usepackage{changepage}
\usepackage{amsmath,amssymb,amsthm,mathrsfs,amsfonts,dsfont,mathtools}
\usepackage[caption=false]{subfig}
\usepackage{epsfig}
\usepackage{color,xcolor}
\usepackage{adjustbox}
\usepackage{tabularx}
\usepackage{array}
\usepackage{multirow}
\usepackage{tabu}
\usepackage{bm}
\usepackage{enumerate}
\usepackage[colorlinks=true,linkcolor=blue,citecolor=blue,urlcolor=blue,pdfauthor={ },pdftitle={ },pdfsubject={ },pdfkeywords={ }]{hyperref}
\usepackage{newtxtext,newtxmath}
\usepackage{makecell} 
\usepackage{url}
\usepackage{booktabs}
\usepackage{nicefrac} 
\usepackage{makeidx}  
\usepackage{leftindex}
\usepackage{algorithm}
\usepackage{algpseudocode}
\usepackage{cleveref}
\usepackage{mwe}

\usepackage{xurl}

\renewcommand\qedsymbol{$\blacksquare$}

\usepackage{changes}

\usepackage{color}
\definecolor{dred}{rgb}{.8,0.2,.2}
\definecolor{ddred}{rgb}{.8,0.5,.5}
\definecolor{dblue}{rgb}{.2,0.2,.8}
\definecolor{dgreen}{rgb}{.2,0.5,.2}

\newtheorem*{theorem*}{Theorem}
\newtheorem*{lemma}{Lemma}
\usepackage{enumitem}

\begin{document}
\title{Measurement-based Dynamical Decoupling for Fidelity Preservation \\on Large-scale Quantum Processors}

\author{Jeongwoo Jae}
\email[JJ: ]{jeongwoo.jae@samsung.com}
\email[CL: ]{changwonlee@yonsei.ac.kr}
\thanks{These authors contributed equally to this work.}
\affiliation{R\&D center, Samsung SDS, Seoul, 05510, Republic of Korea}

\author{Changwon Lee}
\email[JJ: ]{jeongwoo.jae@samsung.com}
\email[CL: ]{changwonlee@yonsei.ac.kr}
\thanks{These authors contributed equally to this work.}
\affiliation{Department of Statistics and Data Science, Yonsei University, Seoul 03722, Republic of Korea}

\author{Juzar Thingna}
\affiliation{American Physical Society, 100 Motor Parkway, Hauppauge, New York 11788, USA}
\affiliation{Center for Theoretical Physics of Complex Systems, Institute for Basic Science (IBS), Daejeon 34126, Republic of Korea}

\author{Yeong-Dae Kwon}
\affiliation{R\&D center, Samsung SDS, Seoul, 05510, Republic of Korea}

\author{Daniel K. Park}
\email{dkd.park@yonsei.ac.kr}
\affiliation{Department of Statistics and Data Science, Yonsei University, Seoul 03722, Republic of Korea}
\affiliation{Department of Applied Statistics, Yonsei University, Seoul 03722, Republic of Korea}
\affiliation{Department of Quantum Information, Yonsei University, Seoul 03722, Republic of Korea}


\begin{abstract}
Dynamical decoupling (DD) is a key technique for suppressing decoherence and preserving the performance of quantum algorithms. We introduce a measurement-based DD (MDD) protocol that determines control unitary gates from partial measurements of noisy subsystems, with measurement overhead scaling linearly with the number of subsystems. We prove that, under local energy relaxation and dephasing noise, MDD achieves the maximum entanglement fidelity attainable by any DD scheme based on bang–bang operations to first order in evolution time. On the IBM Eagle processor, MDD achieved up to a $450$-fold improvement in the success probability of a $14$-qubit quantum Fourier transform, and improved the accuracy of ground-state energy estimation for N$_2$ in the $56$-qubit sample-based quantum diagonalization compared with the standard $XX$-pulse DD. These results establish MDD as a scalable and effective approach for suppressing decoherence in large-scale quantum algorithms.
\end{abstract}

\maketitle



\textit{Introduction---}Decoherence remains one of the central obstacles to realizing full-fledged quantum information processing. Although the theories of quantum error correction (QEC) and fault tolerance are well established~\cite{Shor1996FTQC,Kitaev2003FTQC,Gottesman1998}, the associated resource overhead remains far beyond the capabilities of near-term devices. Developing alternative or complementary strategies to mitigate noise and imperfections is therefore of critical importance. Quantum error mitigation (QEM) has been shown effective in improving the accuracy of expectation-value estimation on current noisy hardware~\cite{Nation2021,Kim2023NatPhys,Kim2023Nat}, where it combines outputs from ensembles of circuits to statistically infer noise-free results, but it remains limited in scope and incurs substantial computational overhead~\cite{Takagi2022,Tsubouchi2023}. In contrast, dynamical decoupling (DD) suppresses decoherence directly through coherent control at the hardware level, offering a more general and scalable means of protecting quantum information that can also be integrated with QEC and QEM techniques.

Originating from nuclear magnetic resonance spectroscopy~\cite{Hahn1950Echo}, DD has since become a widely used method for preserving quantum coherence and now serves as an essential subroutine in high-fidelity quantum computing~\cite{Khodjasteh2002FTQC,Lidar2008FTAQ,Khodjasteh2008Rbound,Ng2011DDQEC,vanderSar2012,Paz-Silva2013}, with demonstrations across trapped ions~\cite{Biercuk2009Ion,Akerman2025Clock}, nitrogen-vacancy centers~\cite{Wang2012NV}, silicon spin qubits~\cite{Jock2022,Park2025}, and superconducting devices~\cite{Ezzell2023,Tripathi2025Qdit,Han2025}. DD reduces noise by applying a sequence of pulses that decouples a system from its environment~\cite{Haeberlen1968AHT}. This sequence typically consists of bang-bang operations, which are strong and nearly instantaneous pulses~\cite{Viola1998two,Viola1999open,Viola1999universal}. The Uhrig DD (UDD) sequence provides near-optimal suppression of single-qubit dephasing~\cite{Uys2009,Biercuk2009Opt,West2010,Uhrig2010OptBound,Pasini2010}, but extending it to multi-qubit systems requires an exponentially increasing number of precisely timed pulses with the system size~\cite{Jiang2011,Wang2011,Kuo2012}, making it difficult to apply to large-scale quantum algorithms. In quantum processors, DD can be implemented at the gate level, where sequences of native gates are applied to idle qubits---those not involved in active logic operations---to suppress decoherence during circuit execution. Recent studies have proposed learning-based strategies to optimize such gate sequences~\cite{Rahman2024,Tong2025}. In particular, a scheme exploiting directed acyclic graph embedding the structure of overlapped idle qubits and qubit connectivity has improved the performance of the quantum Fourier transform and Bernstein–Vazirani algorithms~\cite{Coote2025QCTRL}. However, the computational cost to construct the graph scales quadratically with the number of connected and overlapped qubits, which leads to significant overhead on fully connected architectures. Despite these advances, how additional information can be leveraged to maximize fidelity preservation while minimizing computational cost remains an open challenge.

In this Letter, we introduce measurement-based dynamical decoupling (MDD), a scalable protocol that derives gate sequences from partial measurements. MDD reconstructs the single-qubit reduced density matrix of each idle qubit from Pauli-basis measurements and applies a local unitary that diagonalizes this state, aligning it as closely as possible with its ground state. Although this requires additional measurements, the overhead scales only linearly with the number of idle qubits. We theoretically show that MDD maximally preserves the entanglement fidelity~\cite{Schumacher1996} under local energy relaxation ($T_1$) and dephasing noise ($T_2$), and outperforms all DD strategies based on bang-bang operations. MDD is experimentally validated on the IBM $127$-qubit Eagle processor \textit{ibm\_yonsei}, achieving up to $450$-fold improvement in the success probability of a $14$-qubit quantum Fourier transform~\cite{coppersmith2002QFT} compared to the standard $XX$-pulse DD~\cite{CarrPurcell1954,Meiboom1958MG}. We further demonstrate the scalability and advantage of MDD through its application to the $35$- and $56$-qubit sample-based quantum diagonalization of the N${}_2$ molecule~\cite{Javier2025SQD}, where it outperforms standard DD by yielding higher accuracy and faster convergence in estimating the ground-state energy. These results highlight that leveraging subsystem information enables efficient and robust preservation of quantum algorithm performance against decoherence in large-scale processors.


\begin{figure}[t!]
    \centering
    \includegraphics[width=\linewidth]{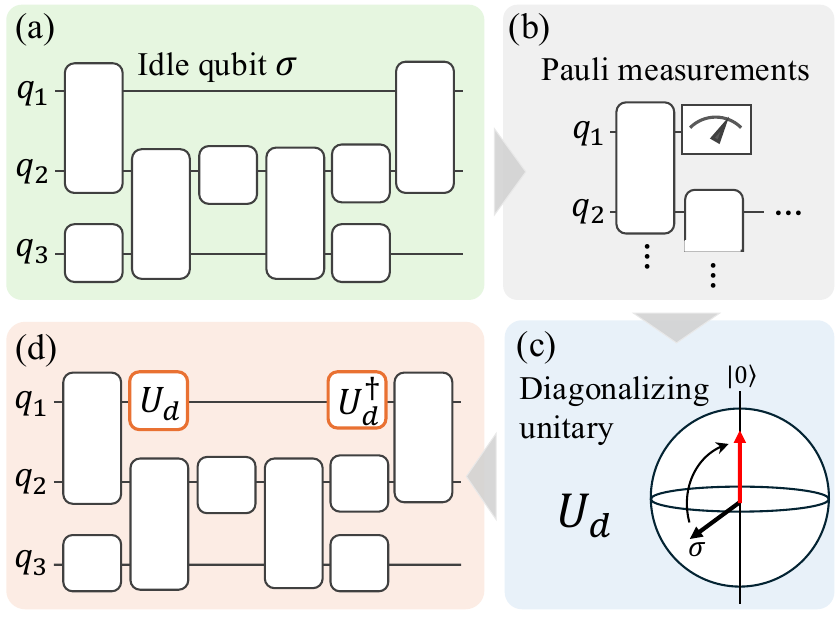}
    \caption{Workflow of MDD. (a) Identify idle qubits of a quantum circuit after transpilation. (b) Perform Pauli-basis measurements on each idle qubit. (c) Classically compute the corresponding diagonalizing unitary $U_d$. (d) Apply the MDD sequence $(U_d,U_d^\dagger)$ to the idle qubit. Steps (b)--(d) are iterated sequentially throughout the quantum circuit over all idle intervals.}
    \label{fig:MDDflow}
\end{figure}

\textit{MDD protocol}---The MDD gate sequence for each idle qubit is defined by a pair of local unitaries $(U_d,U_d^\dagger)$, where $U_d$ diagonalizes the reduced density matrix $\sigma$ of the idle qubit at the beginning of the idle interval, followed by $U_d^\dagger$ at its end. The operation $U_d$ reorders the eigenvalues of $\sigma$ in descending order, i.e., $U_d \sigma U_d^\dagger = \text{diag}(\lambda_1,\lambda_2)$ with $\lambda_1 \ge \lambda_2$, thereby aligning the state as closely as possible with the ground state. This strategy is motivated by the fact that ground states are invariant under typical decoherence processes: for instance, the single-qubit ground state $\ket{0}$ remains unaffected by amplitude damping and dephasing.

Figure~\ref{fig:MDDflow} illustrates the MDD workflow. After circuit transpilation, idle intervals naturally appear between active operations. For each idle qubit, Pauli-basis measurements are performed to estimate its reduced density matrix, from which the corresponding diagonalizing unitary is derived. The resulting MDD sequence $(U_d, U_d^\dagger)$ is then inserted into the circuit to suppress decoherence during idle periods. The expectations of Pauli-basis measurements, $\vec{r}=(\langle X\rangle,\langle Y \rangle,\langle Z \rangle)$, determine the parameters of the diagonalizing unitary operator $U_d = R_y(-\theta_d)R_z(-\phi_d)$ as $\theta_d=\arccos(\langle Z\rangle/r)$ and $\phi_d = \arctan(\langle Y\rangle/\langle X\rangle)$, where $r=\Vert \vec{r}\Vert_2$. Since the rotation about $z$-axis can be implemented virtually by adjusting the phase reference of the control signal, the only physical operation required by MDD is the $R_y$ gate. Requiring only two physical gates per idle interval is also advantageous in practice, as each gate itself is subject to finite duration and imperfections. This process is iterated sequentially across the quantum circuit. For $k$ idle qubits, MDD requires $3k$ Pauli-basis measurements and the classical computation to derive the single-qubit diagonalizing unitaries.

\textit{Optimality of MDD for Fidelity Preservation---}While conventional DD suppresses decoherence by averaging out unwanted system-environment interactions up to a finite order, MDD instead aims to preserve the state fidelity, a quantity directly associated with the performance of quantum algorithms~\cite{Cai2023,KECHEDZHI2024431,Zhou2024coh}. In the following, we show theoretically that, under local decoherence, MDD is the optimal strategy for preserving the fidelity among all DD strategies based on bang–bang operations.

To illustrate this, we consider a pure $N$-qubit state $|\psi\rangle\langle\psi|$, where the $i$-th subsystem is exposed only to local decoherence for a duration $t$. The subsystem is assumed to evolve under a combined amplitude-damping and dephasing channel ${\cal E}$ with relaxation time $T_1$ and dephasing time $T_2$. The ground state $\ket{0}$ serves as the steady state of this channel, toward which any state relaxes. Thus, for a pure single-qubit state, MDD maximally preserves quantum information, as the diagonalized state coincides with the ground state. For a mixed state, the diagonalized state is not identical to the ground state, but MDD fully decouples the state from dephasing noise (see the Supplemental Material (SM)~\cite{SuppMat}).

The optimality of MDD is analyzed against the class of DD protocols that apply $m$ bang-bang operations $g_\alpha \in {\cal S}_g$ at uniform time intervals $\Delta t = t/m$. The preservation of quantum information is quantified by the entanglement fidelity $F_e(|\psi\rangle\langle\psi|,\Lambda) := \langle\psi|\Lambda(|\psi\rangle\langle\psi|) |\psi\rangle$, which measures the overlap between the initial state and its noisy evolution under a quantum channel $\Lambda$~\cite{Schumacher1996,Horodecki1999}. One of our main results is stated as the following theorem:
\begin{theorem*}[\textit{First-order optimality of MDD}]
Let $\Lambda^t_{\mathrm{BB}}$ and $\Lambda^t_{\mathrm{MDD}}$ denote the quantum channels corresponding to a standard DD sequence based on $m$ bang–bang operations $\{g_\alpha\}_{\alpha=1}^m$ and MDD, respectively. Then, for any initial pure state $|\psi\rangle_S$ and sufficiently short evolution time $t$, the entanglement fidelity satisfies $F_e(|\psi\rangle\langle\psi|, \Lambda^t_\mathrm{BB}) \le F_e(|\psi\rangle\langle\psi|, \Lambda_\mathrm{MDD}^{t}) + O(t^2)$.
\end{theorem*}
Therefore, MDD achieves the maximum entanglement fidelity up to first order in the evolution time $t$ among all DD protocols based on bang–bang operations.

To prove this result, we first establish the following lemma: the MDD sequence $(U_d, U_d^\dagger)$ maximizes the entanglement fidelity among all pairs of single-qubit unitaries and their adjoints $(U,U^\dagger)$, i.e.,
\begin{eqnarray}
    F_e\big(|\psi\rangle\langle\psi|,\Lambda^t_{U_d}\big) =\max_U F_e\big(|\psi\rangle\langle\psi|, \Lambda^t_{U}\big),~\forall t,
\end{eqnarray}
where $\Lambda^t_{U}:=({\cal U}^\dagger \circ{\cal E}^t\circ{\cal U})_i$ denotes the quantum channel obtained by conjugating the decoherence process ${\cal E}^t_i$ with the unitary channel ${\cal U}_i$, both acting locally on the $i$-th subsystem.

The theorem follows from the observation that, for the same evolution time, the concatenation of the optimal strategies yields the entanglement fidelity that cannot be exceeded by any concatenation of sub-optimal strategies. Specifically, applying the bang-bang operations at time intervals $\Delta t$, the system evolves under a channel $\Lambda_\text{BB}^t$ defined in the toggling frames of the cumulative control operations $U_\alpha = g_\alpha g_{\alpha-1}\cdots g_1$ satisfying $U_{\alpha} U^\dagger_{\alpha-1} = g_\alpha$ as $\Lambda_\text{BB}^t = \Lambda^{\Delta t}_{U_m}\circ \cdots \circ\Lambda^{\Delta t}_{U_1}$. To first order in the evolution time, the entanglement fidelity of $\Lambda_\text{BB}^t$ can be approximated as
\begin{eqnarray}
\label{eq:BBapprox}
F_e(|\psi\rangle\langle\psi|,\Lambda^t_\text{BB})\approx \frac{1}{m}\sum_{\alpha=1}^m F_e(|\psi\rangle\langle\psi|, \Lambda^t_{U_\alpha}) + O(t^2).
\end{eqnarray}
From the lemma, each term in the summation satisfies $F_e(|\psi\rangle\langle\psi|, \Lambda^t_{U_\alpha}) \le F_e(|\psi\rangle\langle\psi|, \Lambda^t_{U_d})$ for all $\alpha$. Hence, the fidelity attained by any bang-bang sequence cannot exceed that of sequentially applying $m$ MDD operations at the same intervals $\Delta t$. Since $U_d U_d^\dagger=I$, concatenating $m$ MDD operations is equivalent to a single application over the entire duration $t$, i.e., $\Lambda_{U_d}^{\Delta t}\circ\cdots\circ\Lambda_{U_d}^{\Delta t} = \Lambda_{U_d}^{t}$. This yields $F_e(|\psi\rangle\langle\psi|, \Lambda^t_\text{BB}) \le F_e(|\psi\rangle\langle\psi|, \Lambda_{U_d}^{t}) + O(t^2)$, which constitutes the theorem. The detailed derivation, together with its generalization to multiple subsystems under uncorrelated decoherence, is presented in the SM~\cite{SuppMat}.

\begin{figure}[t!]
    \centering
    \includegraphics[width=\linewidth]{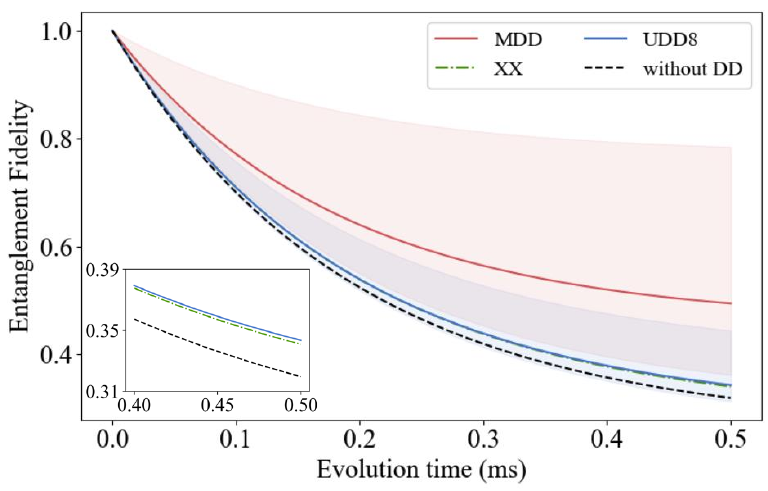}
    \caption{Entanglement fidelity of four-qubit states ($N=4$) under different DD schemes: MDD (red), $XX$ (green), UDD$8$ (blue), and without DD (black). Each curve indicates the average over the same $20$ randomly chosen states. The red (MDD) and blue (UDD) shaded regions represent the maximum and minimum values of fidelity across these states. The relaxation and dephasing times are set to $T_1 = 250\mu s$ and $T_2=170\mu s$, respectively, matching typical values of the IBM Eagle processor.}
    \label{fig:vsUDD}
\end{figure}

Figure~\ref{fig:vsUDD} shows that MDD outperforms $XX$ and UDD for $N=4$. The $XX$ protocol applies $X$ gates at $25\%$ and $75\%$ of idle interval~\cite{CarrPurcell1954,Meiboom1958MG}. A UDD$n$ sequence can suppress pure dephasing of a single qubit up to order ${n}$ by applying $\pi$ pulses at nonuniform times $t_\alpha = t\sin^2(\alpha\pi/(2n+2))$~\cite{Uhrig2007UDD,Lee2008MinDP}. These nonuniform intervals can effectively be interpreted as a uniform sequence of bang-bang operations including the identity operations; hence, by the theorem, one can anticipate that MDD outperforms UDD under the local decoherence. Further comparisons with standard DD protocols, including $XY4$~\cite{MAUDSLEY1986488}, and quadratic DD (QDD)~\cite{West2010,Kuo2011QDD,Quiroz2011QDD}, are provided in the SM~\cite{SuppMat}. Although UDD and QDD are regarded as among the most effective strategies for suppressing decoherence in superconducting qubit systems~\cite{Ezzell2023}, our simulations show that MDD achieves higher entanglement fidelities than all conventional DD methods.

We further analyze the optimality of MDD in terms of the decay rate of the entanglement fidelity, $\lvert dF_e/dt \rvert$. As detailed in the SM~\cite{SuppMat}, the MDD sequence $(U_d,U_d^\dagger)$ minimizes this decay rate among all pairs of single-qubit unitaries $(U,U^\dagger)$. Under local relaxation and dephasing noise acting on $i$-th subsystem, described by Lindblad operators ${L}_1 = (\sigma_{x}+i\sigma_y)_i/2$ and $L_2=\sigma_{z,i}$, respectively, the decay rate satisfies $\vert \langle\psi|d\Lambda^t_U(|\psi\rangle\langle\psi|)/dt |\psi\rangle\vert = \sum_k \Gamma_k \text{Var}_{\sigma_U}[L_k]$, where $\text{Var}_\sigma[A] := \Tr(\sigma A^\dagger A) - \vert\Tr(\sigma A)\vert^2$, $\sigma_U = U\sigma U^\dagger$, $\Gamma_1 = 1/T_1$, and $\Gamma_2 = 1/(T_2 - 1/2T_1)$. This value is minimized when $U=U_d$, indicating that MDD effectively delays the passage time over which a quantum state evolves toward its ground state. We also extend the analysis, as shown in the SM~\cite{SuppMat}, to a two-qubit MDD scheme tailored to correlated decoherence arising from $ZZ$ crosstalk between subsystems $i$ and $j$~\cite{Sundaresan2020,Ku2020,Baum2021DRL,Coote2025QCTRL}. For a pure state $\ket{\psi}_{ij}$, the MDD sequence can be defined by a two-qubit unitary $U_{ij}$ such that $U_{ij}\ket{\psi}_{ij} = \ket{00}_{ij}$ since the ground state $\ket{00}_{ij}$ is a steady-state under relaxation, dephasing, and $ZZ$ crosstalk. Implementing such two-qubit MDD requires entangling operations that may introduce additional idle intervals elsewhere in the circuit. Therefore, we adopt the single-qubit MDD as a practical heuristic approach within the scope of this work.

\textit{Experiments}---To demonstrate the scalability and effectiveness of MDD, we evaluate its performance using two representative quantum algorithms: the Quantum Fourier Transform (QFT) and the sample-based quantum diagonalization (SQD). QFT is a key subroutine underlying many quantum algorithms that exhibit quantum advantage~\cite{coppersmith2002QFT}, while SQD is a quantum-classical hybrid framework for estimating molecular ground-state energies and is considered a promising approach for showcasing practical quantum utility~\cite{Javier2025SQD}. MDD is compared with the standard $XX$ DD scheme and the baseline without DD. Each MDD sequence is defined using $10^4$ measurement shots per Pauli observable. Experiments are conducted on the IBM $127$-qubit Eagle processor \textit{ibm\_yonsei}, with device specifications provided in the SM~\cite{SuppMat}.

\begin{figure}[b!] 
    \centering
    \includegraphics[width=\linewidth]{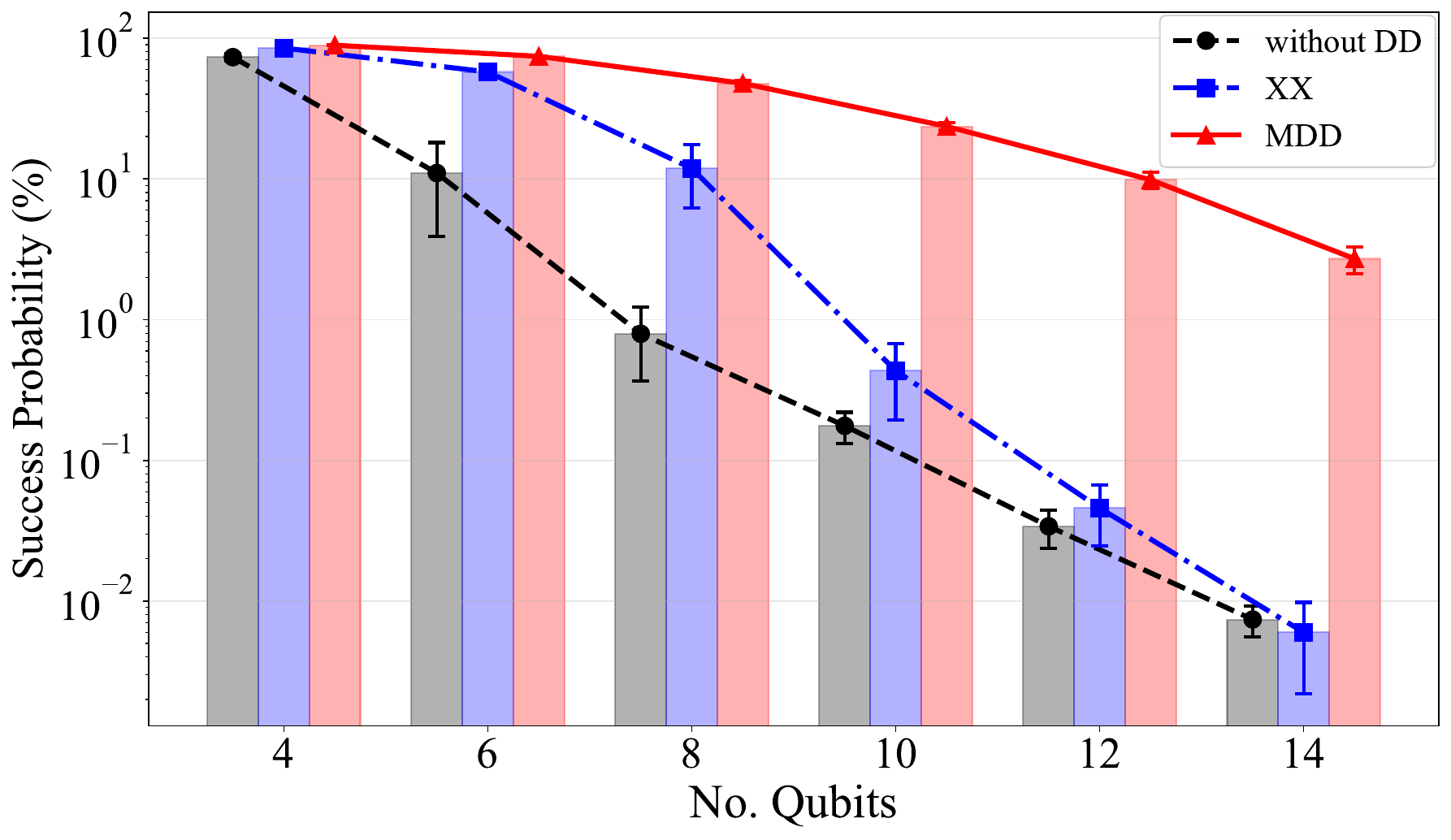}
    \caption{Success probability of QFT circuits up to $14$ qubits. Without DD (black dashed line), the success probability rapidly decreases as the number of qubits increases, implying the accumulation of errors. MDD (red solid line) preserves the QFT performance significantly better than the standard $XX$ scheme (blue dot-dashed line). Each point represents the mean success probability over five independent experiments, and error bars indicate the standard deviations.
    }
    \label{fig:fig_qft}
\end{figure}

QFT circuits consist of successive controlled-phase gates that cause qubits acted on early in the sequence to remain idle for long periods. Even after transpilation, a substantial number of idle qubits remain (see the SM~\cite{SuppMat}), leading to significant error accumulation as circuit size increases.

The results of the QFT experiments are shown in Fig.~\ref{fig:fig_qft}. We use the success probability as the performance metric, defined as
$
    p_\text{success} :=  n_\text{target}/n_\text{shots}\times 100 \%, 
$
where $n_\text{target}$ is the number of occurrences of the target output bit string and $n_\text{shots}$ is the total number of measurements. The input state is prepared in the computational basis, for which the ideal QFT output (target) corresponds to the alternating $N$-bit string $0101\ldots01$~\cite{Mundada2023QCTRL}. Each experiment uses $n_\text{shots}=10^{5}$ and repeated five times. While the baseline without DD and the standard $XX$ scheme exhibit a rapid decay in success probability as the number of qubits $N$ increases, the performance of MDD remains substantially higher. For $N=10$, MDD achieves about a $50$-fold improvement over $XX$, and for $N=14$, nearly $450$-fold, demonstrating a clear advantage against conventional DD in protecting idle qubits in deep circuits.

SQD is a quantum-classical hybrid scheme to estimate the energy of molecules. It begins by simulating electronic interaction of a molecule using the Local Unitary Cluster Jastrow (LUCJ) ansatz~\cite{motta2023bridging}, followed by acquisition of noisy measurement samples $\tilde{X}$ from the quantum device. These samples are refined classically through self-consistent configuration recovery, which identifies configurations with incorrect particle numbers and probabilistically corrects them using an iteratively updated average spin–orbital occupancy vector $\mathbf{n}$. The corrected configuration set is then used to form subspaces in which the molecular Hamiltonian is diagonalized on a classical computer to obtain the ground-state energy. A detailed description of the SQD algorithm is provided in the SM~\cite{SuppMat}. Thus, suppressing noise can improve the quality of samples and consequently, the effectiveness of the self-consistent configuration recovery process. Since SQD relies on sampling rather than estimation of expectation values, conventional quantum error mitigation techniques such as zero-noise extrapolation and probabilistic error amplification cannot be straightforwardly applied~\cite{Nation2021,Kim2023NatPhys,Kim2023Nat}.
In contrast, DD operates at the hardware level to suppress decoherence during circuit execution, making it a natural and scalable approach for improving the ground-state estimation of SQD.

\begin{figure}[t!]
    \centering
    \includegraphics[width=\linewidth]{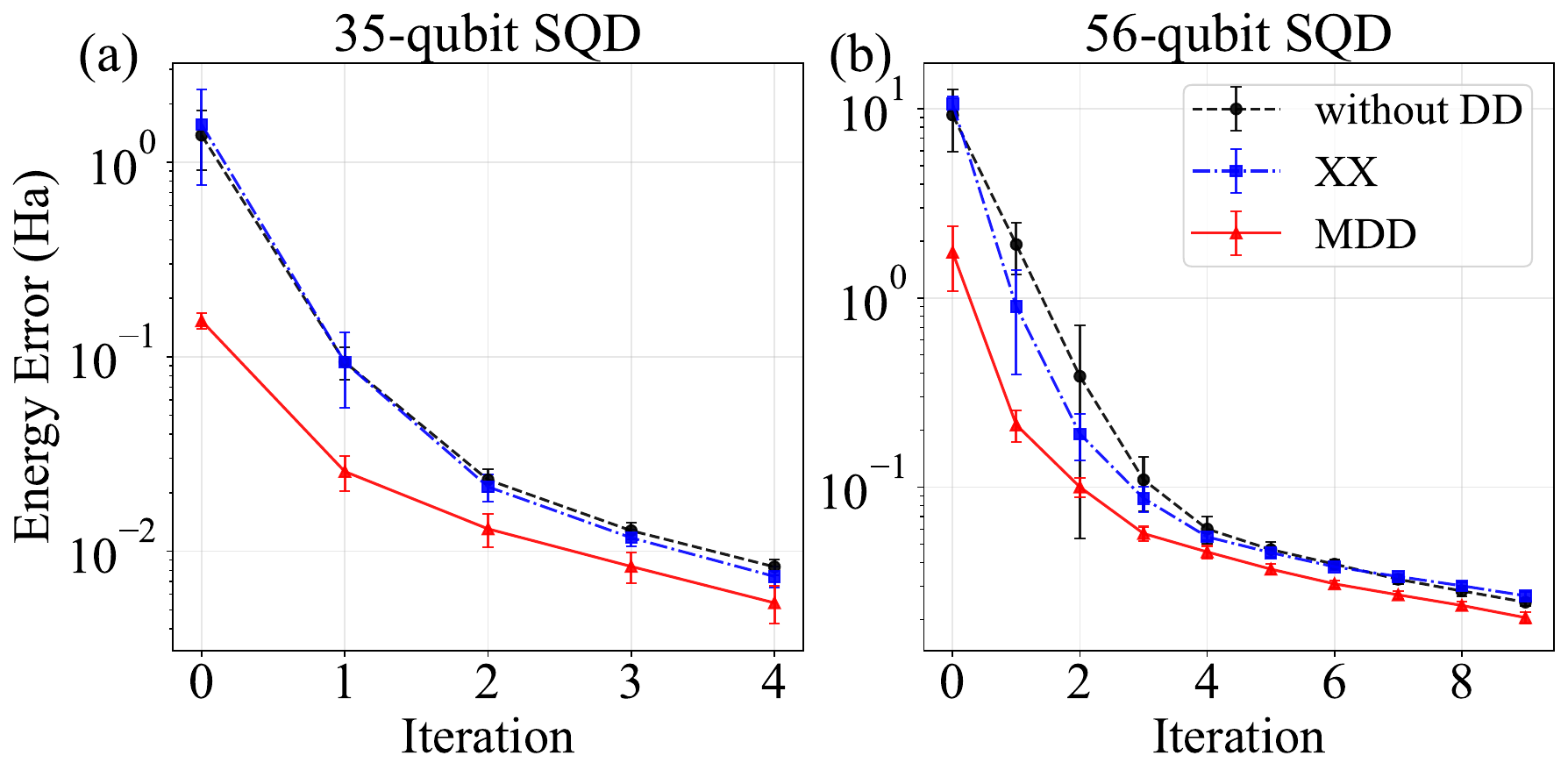}
    \caption{Comparison of estimated ground-state energy errors for $\mathrm{N}_2$ obtained using the SQD workflow with three DD schemes: without DD (black dashed line), $XX$ (blue dotdashed line), and MDD (red solid line). Panels (a) and (b) show the mean energy errors versus the number of self-consistent configuration recovery iterations for $35$- and $56$-qubit experiments, respectively. Error bars represent the standard deviation over five independent runs.
    }
    \label{fig:fig_sqd}
\end{figure}

We apply SQD to estimate the ground-state energy of the nitrogen molecule ($\mathrm{N}_2$) using $35$- and $56$-qubit circuits with the $6$-$31$G and cc-pVDZ basis sets, respectively. The LUCJ gate parameters are initialized from the classical coupled cluster singles and doubles (CCSD) method~\cite{sun2018pyscf}. The performance of each DD method is evaluated by comparing the estimated energy with the exact energy. Each experiment employs $10^5$ shots for the $35$-qubit case and $10^7$ shots for the $56$-qubit case, with five and ten self-consistent configuration recovery iterations, respectively.

Experimental results for the ground-state energy errors as a function of the number of recovery iterations are shown in Fig.~\ref{fig:fig_sqd} (a) and (b) for $35$- and $56$-qubit cases, respectively. MDD consistently yields lower initial energy errors and faster convergence during the self-consistent configuration recovery iterations. In the $35$-qubit experiment, MDD reduces the initial energy error to $0.15 \pm 0.01$ Ha, approximately $9$ and $10$ times lower than those obtained without DD and with the $XX$ sequence, respectively. In the $56$-qubit experiment, MDD achieves an initial error of $1.75 \pm 0.66$ Ha, about $5.3$ and $6.1$ times lower than those obtained without DD and with the $XX$ sequence, respectively. These results clearly demonstrate that MDD effectively preserves the performance of large-scale quantum chemistry algorithms by suppressing decoherence.



The high-quality samples produced by MDD further improve the outcomes of the self-consistent configuration recovery procedure. After the iterative recovery process, MDD yields the lowest final energy errors. In the $35$-qubit experiment, the error converges to $5.4 \times 10^{-3} \pm 1.2 \times 10^{-3}$ Ha after five iterations, corresponding to an accuracy approximately $1.4$ times higher than that of $XX$. Similarly, in the $56$-qubit experiment, MDD achieves a final energy error of $2.05 \times 10^{-2} \pm 1.5 \times 10^{-3}$ Ha after ten iterations, about $1.3$ times more accurate than $XX$. These results indicate that MDD enhances the convergence of classical postprocessing, thereby improving the overall efficiency of the SQD workflow. Notably, as shown in the SM~\cite{SuppMat}, MDD enables the \textit{ibm\_yonsei} (Eagle) processor to attain nearly the same accuracy as the more advanced \textit{ibm\_fez} (Heron) device. Although Heron, with its shorter coherence times, exhibits smaller initial energy errors, both processors reach comparable accuracies after several recovery iterations, underscoring MDD’s ability to extract near–state-of-the-art performance even from less advanced hardware.

\textit{Conclusions---}We have introduced MDD, a scalable protocol that preserves the performance of quantum algorithms by optimally protecting idle qubits from local decoherence. Experiments on the $127$-qubit IBM Eagle processor demonstrate up to a $450$-fold performance enhancement in a $14$-qubit QFT, and significantly improved accuracy in the energy estimation of the N$_2$ molecule. These results establish MDD as an efficient and hardware-compatible approach for suppressing noise in large-scale quantum algorithms.

\textit{Acknowledgments.---}J.J. and Y.-D. K. thank Taehee Lee and Youngjune Gwon for support. This work is supported by the Samsung SDS Industry-Academia Collaboration Program.


%


\clearpage
\widetext
\begin{center}
\textbf{\large Supplemental Material: Measurement-based Dynamical Decoupling for Fidelity Preservation \\on Large-scale Quantum Processors}
\end{center}

\setcounter{equation}{0}
\setcounter{section}{0}
\setcounter{table}{0}
\setcounter{figure}{0}
\renewcommand{\thetable}{S\arabic{table}}
\renewcommand{\d}[1]{\ensuremath{\operatorname{d}\!{#1}}}
\renewcommand{\thesection}{S\arabic{section}}
\renewcommand{\thesubsection}{S\arabic{section}-\arabic{subsection}}
\renewcommand{\theequation}{S\arabic{equation}}
\renewcommand{\thetable}{S\arabic{table}}
\renewcommand{\thefigure}{S\arabic{figure}}

\section{Decoherence model}
The dynamics of local decoherence in quantum devices can be modeled using the combination of amplitude damping channel $\{K^\text{AD}_1,K^\text{AD}_2\}$ and dephasing channel $\{K^\text{DP}_1,K^\text{DP}_2\}$, where Kraus operators are represented by
\begin{eqnarray}
\label{eq:Kraus}
    K^\text{AD}_1 &=& aI+bZ,\qquad K^\text{AD}_2=\frac{\sqrt{\gamma_1}}{2}(X+iY) \nonumber\\
    K^\text{DP}_1 &=& \sqrt{\alpha}I,\quad\qquad K^\text{DP}_2=\sqrt{\beta}Z,
\end{eqnarray}
for $a={(1+\sqrt{1-\gamma_1})}/{2}$, $b={(1-\sqrt{1-\gamma_1})}/{2}$, $\alpha={(1+\gamma_p)}/{2}$, $\beta={(1-\gamma_p)}/{2}$, and $\gamma_p = \gamma_2/\sqrt{\gamma_1}$. $\gamma_p = e^{-t/T_p} \in[0,1]$ for $1/T_p = 1/T_2 - 1/2T_1$. These operators satisfy $\sum_i  K^{\text{AD}\dagger}_i K^\text{AD}_i=I$ and $\sum_j K^{\text{DP}\dagger}_j K^\text{DP}_j =I$. The combined decoherence channel $\cal E$ is obtained by a set of the combined Kraus operators $\{M_{ij}=K^\text{DP}_iK^\text{AD}_j\}$, where $M_{11}=\sqrt{\alpha}K^\text{AD}_1$, $M_{12}=\sqrt{\alpha}K^\text{AD}_2$, $M_{21}=\sqrt{\beta}(bI+aZ)$, and $M_{22}=\sqrt{\beta}K^\text{AD}_2$. By this channel, an input single-qubit state $\varrho$ is transformed as
\begin{eqnarray}
\label{eq:combined}
    {\cal E}^t(\varrho) := \sum_{i,j=0}^1 M_{ij}\varrho M^\dagger_{ij}=\begin{pmatrix}
        \varrho_{00} + (1-\gamma_1)\varrho_{11} & \gamma_2\varrho_{01}  \\
        \gamma_2 \varrho_{10} & \gamma_1\varrho_{11}
    \end{pmatrix}.
\end{eqnarray}
We will consider decoherence involving $ZZ$ crosstalk later.

For a general mixed state, MDD decouples the state from the dephasing as the diagonalized state $\sigma_d = (I+\eta Z)/2$ for $0\le \eta < 1$ commutes with the Kraus operators of dephasing, i.e., $[K^\text{DP}_j,\sigma_d]=0$ for all $j$, and the state is invariant under the dephasing $\sum_j K^\text{DP}_j \sigma_d K^{\text{DP}\dagger}_j = \sigma_d$. When a state is pure ($\eta =1$), MDD aligns the state with the ground state $|0\rangle\langle 0| = (I+Z)/2$. The ground state is invariant under the combined decoherence~\eqref{eq:combined}: ${\cal E}^t(|0\rangle\langle 0|) =  \sum_{i,j=0}^1 M_{ij}|0\rangle\langle 0| M^\dagger_{ij} = |0\rangle\langle 0|$.

\section{Proofs of the Lemma and Theorem}

We first prove the lemma which states that the MDD is an optimal process to achieve the maximum entanglement fidelity over any process employing a pair of conjugate unitary operators applied at the beginning and end of the decoherence time. Based on the analysis used for the lemma, we also investigate the fidelity change induced by a gate error, and the bounds of the entanglement fidelity for mixed states. We introduce an approximate representation of the entanglement fidelity of DD using the bang-bang operations. Using the lemma and the approximate representation of the bang-bang operations, we prove the theorem. Furthermore, we generalize the theorem to a scenario of multiple subsystems under uncorrelated decoherence. Finally, we numerically compare MDD with $XX$, $XY4$, UDD, QDD and a new scheme which combines the MDD and the $XX$. For two-qubit states, the comparison with UDD is also performed under Ohmic and $1/f$ noise induced by random phase noise.

\subsection{The optimality of the diagonalizing unitary operator}
We represent the action of the MDD and the decoherence channel as $\Lambda^t_\text{MDD}={\cal U}_d^\dagger\circ{\cal E}^t\circ{\cal U}_d$, where ${\cal U}_d(\cdot) = U_{d,i}\otimes I_{S/i} (\cdot) U_{d,i}^\dagger\otimes I_{S/i}$ is a unitary channel and ${\cal E}^t={\cal E}^t_i\otimes {\cal I}_{S/i}$ is the decoherence channel, both acting on the $i$-th subsystem of the total system $S$. The performance of DD can be quantified by the fidelity between the original $N$-qubit state $|\psi\rangle\langle \psi|_S$ and the dynamically decoupled state $\Lambda^t_\text{MDD}(|\psi\rangle\langle \psi|_S)$ as
\begin{eqnarray}
\label{eq:fidelity}
    F\big(|\psi\rangle\langle \psi|_S,\Lambda^t_\text{MDD}\left(|\psi\rangle\langle \psi|_S\right)\big),
\end{eqnarray}
where $F(X,Y):=(\Tr\sqrt{\sqrt{X}Y\sqrt{X}})^2$. This quantity is also called entanglement fidelity, denoted by $F_e\left(|\psi\rangle\langle \psi|_S, \Lambda^t_\text{MDD}\right)$~\cite{Schumacher1996}. (Note that we use $\Lambda^t_\text{MDD}$ and $\Lambda^t_{U_d}$ interchangeably.) We compare the MDD with a DD replacing $U_{d,i}$ and $U_{d,i}^\dagger$ with an arbitrary unitary operator $U_{i}$ and its self-adjoint operator $U_{i}^\dagger$, respectively. We represent the action of this DD as a channel $\Lambda^t_U={\cal U^\dagger}\circ {\cal E}^t \circ {\cal U}$.

For the entanglement fidelities of the processes $\Lambda_\text{MDD}$ and $\Lambda_U$, the following holds:
\begin{lemma}
    $\Lambda_\text{MDD}$ maximizes the entanglement fidelity among all processes $\Lambda_U$, i.e.,
    \begin{equation}
    F_e(|\psi\rangle\langle\psi|_S,\Lambda^t_\text{MDD}) = \max_U F_e(|\psi\rangle\langle\psi|_S,\Lambda^t_U),~\forall t.
    \end{equation}
\end{lemma}
This implies that $\Lambda^t_\text{MDD}$ is optimal in preserving the entanglement between the $i$-th subsystem and the remaining part.

{\em Proof of Lemma.---}Since the original state is pure, the entanglement fidelity $F_e(|\psi\rangle\langle\psi|_S,\Lambda_U)$ can be written in terms of the reduced density matrix $\sigma_i$ of the $i$-th single-qubit subsystem, the unitary operators $(U_{i},U_{i}^\dagger)$ for the dynamical decoupling, and the Kraus operators $\{M_{jk}\}$ of the decoherence channel:
\begin{eqnarray}
    F_e\big(|\psi\rangle\langle\psi|_S , \Lambda^t_U\big) = \sum_{j,k}\left\vert{\Tr\left(M_{jk}U_i\sigma_i U_i^\dagger\right)}\right\vert^2.
\end{eqnarray}
According to the parameterization in~\eqref{eq:Kraus}, the right-hand side can be represented by a quadratic function $f$ as
\begin{eqnarray}
\label{eq:G}
    f(r_z) = \alpha\left(a+br_z\right)^2 + \beta\left(b+ar_z\right)^2 + \frac{\gamma_1}{4}\left(r^2-r_z^2\right),
\end{eqnarray}
where $r_z\in[-r,r]$ is the $z$ component of the Bloch vector $\vec{r}$ of the reduced density matrix $U\sigma U^\dagger$ satisfying $\Vert\vec{r}\Vert=r\in[0,1]$.

Let the extreme point of $f(r_z)$ be $r_z^\star$, then, depending on the sign of the second derivative, we have the following three cases:
\begin{enumerate}[label={(C\arabic*)}]

    \item $f''(r_z)>0:\quad r_z^\star \le 0$.
    
    \item $f''(r_z)<0:\quad r_z^\star \ge 1$.
    
    \item $f''(r_z)=0:\quad f'(r_z)\ge 0,~\forall r_z$.
    
\end{enumerate}
Since the function is positive semidefinite, the maximum value of $f(r_z)$ can be achieved at the point $r_z = r$ for all three cases: Let $\sqrt{1-\gamma_1}=s\in[0,1]$. The second derivative of the function is $f''(r_z)=s(s-\gamma_p)$. By the first derivative, the extreme point is given by $r_z^\star = -2ab/f''(r_z)$, where $2ab = (1-s^2)/2$ is positive semidefinite. If $f''(r_z)>0$ and the function is convex, $r^\star_z\le0$ and $f$ has the maximum value at $r_z=r$. (Note that, if $s=0$ and $r^\star_z = 0$, the decoherence model becomes the daphasing channel, and the maximum values appear at two points $r=-r$ and $r$.) If $f''(r_z)<0$ and the function is concave, we can deduce the location of the extreme point from $D = f''(r_z) + 2ab$. $D=(s^2-2\gamma_p s+1)/2$ and this is a positive semidefinite function as $\gamma_p \in [0,1]$ and $1-\gamma_p^2 \ge 0 $. $D\ge0$ implies that the extreme point $r^\star_z = -2ab/f''(r_z)\ge1$ and $f$ is an increasing function in its domain $[-r,r]$. Thus, the maximum value can be achieved at $r_z=r$. If $f''(r_z)=0$, $f$ becomes a linear function of which slope is positive semidefinite by $f'(r_z)=2ab$ and the function increases monotonically in the domain $[-r,r]$. This implies that the maximum value appears at $r_z=r$.

In other words, the maximum value of the function is attained at the maximal element of its domain, which is the point $r_z=r$, ($r_x=0$ and $r_y=0$). This can be achieved by the unitary operator $U = U_d$, which diagonalizes the reduced density matrix $\sigma_i$ of the original state and aligns its eigenvalues in descending order. Thus, $\Lambda_\text{MDD}$ maximizes the entanglement fidelity among all processes $\Lambda_U$. \hfill\qedsymbol

The maximum value of entanglement fidelity is $f(r)=\alpha(a+br)^2 + \beta(b+ar)^2$; if the noisy subsystem is pure ($r=1$), $f(1) = 1$; if the reduced density matrix represents the maximally mixed state ($r=0$), $f(0)=\alpha a^2 + \beta b^2$.

\subsection{Gate error and Bounds of fidelity for mixed states}

We further investigate the fidelity change induced by a gate error and the bounds of entanglement fidelity for mixed states. Assume that the unitary operation has a gate error $\delta$, which causes the operation $U_d$ to be tilted and the $z$ component of the reduced density matrix becomes $r_\delta=r\cos\delta$. Then, the amount of fidelity change caused by the gate error becomes
\begin{eqnarray}
    f(r) - f(r_\delta) &\approx& \frac{r\delta^2}{4}\left[(1-2r)\gamma_1 +2r\left(1-\gamma_p\sqrt{1-\gamma_1}\right) \right] + {\cal O}(\delta^4).
\end{eqnarray}
Thus, the fidelity is decreased by the gate error.

Since the fidelity is never decreased by a non-selective quantum operation and invariant under a unitary operation, the fidelity between the original input mixed state $\varrho_S$ and the dynamical decoupled state $\Lambda^t_\text{MDD}(\varrho_S)$ is bounded by
\begin{eqnarray}
    F_e\left( \sigma_d, {\cal E}_i\right)
    \ge F_e\left(\varrho_S, \Lambda^t_\text{MDD}\right)\ge F_e\left(|\phi\rangle\langle\phi| , \Lambda^t_\text{MDD}\right),
\end{eqnarray}
where $\sigma_d = U_d\sigma_i U^\dagger_d=\text{diag}(\lambda_1,\lambda_2)$ is the diagonalized density matrix of the subsystem $\sigma_i = \Tr_{S/i}(\varrho_S)$ and the pure state $|\phi\rangle\langle\phi|$ is a purification of the input mixed state $\varrho_S$. We have analytic expression of the bounds:
\begin{eqnarray}
    F_e\left( \sigma_d, {\cal E}_i\right) = \Tr(\sigma_d {\cal E}_i(\sigma_d))+ 2\sqrt{\det(\sigma_d)\det({\cal E}_i(\sigma_d))},\nonumber
\end{eqnarray}
and
\begin{eqnarray}
    F_e\left(|\phi\rangle\langle\phi| , \Lambda^t_\text{MDD}\right) = \sum_{i,j}\left\vert{\Tr\left(M_{ij}\sigma_d\right)}\right\vert^2.
\end{eqnarray}
If the subsystem is in a pure state, $F\left(\varrho_S,\Lambda^t_\text{MDD}(\varrho)\right)=1$ as $F_e\left( \sigma_d, {\cal E}_i\right)=F_e\left(|\phi\rangle\langle\phi|, \Lambda^t_\text{MDD}\right)=1$. For the maximally mixed state $\sigma_d = I/2$, $F_e\left( \sigma_d, {\cal E}_i\right)=\left(1+\sqrt{1-\gamma_1^2}\right)/2$ and $F_e\left(|\phi\rangle\langle\phi| , \Lambda^t_\text{MDD}\right)=(2-\gamma_1+2\gamma_p \sqrt{1-\gamma_1})/4$. This derivation shows that the reduced density matrix determines the both bounds of the entanglement fidelity and this underscores the importance of the information of noisy subsystem in preserving entanglement from decoherence.

\subsection{First-order representation of the typical DD and the entanglement fidelity}
We describe the typical DD based on the bang-bang operations with open quantum systems. Consider that a system $S$ interacts with a bath $B$, and its Hamiltonian is given by $H_0 = H_S\otimes I_B + I_S\otimes H_B + H_{SB}$, where $H_s$ is the Hamiltonian of the system, $H_B$ is the Hamiltonian of bath, and $H_{SB}$ is the interaction Hamiltonian. DD can be implemented with control pulses described by a time-dependent Hamiltonian $H_{S,c}(t)$, so the total Hamiltonian becomes $H_0(t)=H_0 + H_{S,c}(t)$. We assume that the typical DD applies a control unitary operator $g_\alpha  \in {\cal S}_g$ at time $\alpha\Delta t$, where $\vert{\cal S}_g\vert=m$ and $\Delta t = t/m$. The control Hamiltonian is given by
\begin{eqnarray}
    H_{s,c}(t) = \sum_{\alpha=1}^m \delta(t-\alpha\Delta t) H_{c,\alpha}.
\end{eqnarray}
We consider that the pulse width is negligible, so the unitary of a bang-bang operation becomes $g_\alpha = \exp(-it H_{c,\alpha})$.

In this setting, the initial state of system $\varrho_S$ evolves according to the channel $\Lambda_\text{BB}^t$ defined by
\begin{eqnarray}
    \Lambda_\text{BB}^t(\varrho_S) := {\cal G}_m\circ{\cal E}_m^{\Delta t}\cdots{\cal G}_2\circ{\cal E}_2^{\Delta t}\circ {\cal G}_1\circ{\cal E}_1^{\Delta t}(\varrho_S),
\end{eqnarray}
where ${\cal G}_\alpha(\varrho_S) = g_\alpha\varrho_S g_\alpha^\dagger$. Here, we assume the stroboscopic stationarity; the local decoherence channel ${\cal E}^{\Delta t}_\alpha$ defined in~\eqref{eq:combined} is considered identical in all time intervals. A sequence of symmetrized bang-bang operations can be obtained by setting ${\cal G}_m$ to the identity channel ${\cal I}$. Now we define the control unitary operator of $H_{S,c}(t)$:
\begin{eqnarray}
    U_{\alpha} := {\cal T}\!\!\exp\left(-i\int_0^{\alpha\Delta t} H_{S,c}(t) dt \right) = {\cal T}\!\!\prod_{t=1}^\alpha g_t.
\end{eqnarray}
The control unitary operator can be represented by the product of bang-bang operations and, at the end of time evolution, it becomes the identity up to a global phase $\varphi$ as $U_{m} = g_m g_{m-1}\ldots g_1 = e^{i\varphi}I$. Since $U^\dagger_{\alpha-1} U_{\alpha} = g_\alpha$, the channel of the bang-bang operations can be written in the toggling frame as
\begin{eqnarray}
    \Lambda^t_\text{BB}(\varrho_S) = \Lambda^{\Delta t}_{U_m}\circ\Lambda^{\Delta t}_{U_{m-1}}\circ\cdots\circ\Lambda^{\Delta t}_{U_1}(\varrho_S).
\end{eqnarray}

In the first-order approximation of each channel ${\cal E}_\alpha(\cdot)\approx {\cal I} + {\cal L}_\alpha(\cdot) \Delta t$ for the superoperator ${\cal L}_\alpha$ satisfying ${\cal L}_\alpha = {\cal L}_\beta={\cal L}$ $\forall \alpha,\beta\in[m]$, a quantum state of the system $\varrho_S$ after the channel $\Lambda^t_\text{BB}$ can be written as
\begin{eqnarray}
    \Lambda^t_\text{BB}(\varrho_S) &\approx& {\cal I}+\sum_{\alpha=1}^m U_\alpha^\dagger{\cal L}_\alpha (U_\alpha  \varrho_S U_\alpha ^\dagger)U_\alpha \Delta t + O(\Delta t^2) \nonumber\\
    &\approx&\frac{1}{m}\sum_{\alpha=1}^m U_\alpha^\dagger {\cal E}^t \left(U_\alpha  \varrho_S U_\alpha ^\dagger\right)U_\alpha  + O(t^2) \nonumber\\
    &=&\frac{1}{m}\sum_{\alpha=1}^m\Lambda^t_{U_\alpha }(\varrho_S) + O(t^2),
\end{eqnarray}
where ${\cal E}^t(\varrho_S) \approx {\cal I}+{\cal L}(\varrho_S)t$ and $\Delta t = t/m$ for a finite $m$. This holds when the first and second derivatives of the density matrix ${\cal E}^t(\varrho_S)$ are finite.

Based on this, for a pure state $\varrho_S=|\psi\rangle\langle\psi|_S$, the entanglement fidelity can then be written as
\begin{eqnarray}
\label{eq:fidelity_approx}
    F_e(|\psi\rangle\langle\psi|_S, \Lambda_\text{BB}) &\approx& \frac{1}{m}\sum_{\alpha=1}^m F_e\big(|\psi\rangle\langle\psi|_S, \Lambda^t_{U_\alpha }\big) + O(t^2)\nonumber\\
    &=& F_e\bigg(|\psi\rangle\langle\psi|_S, \frac{1}{m}\sum_{\alpha=1}^m\Lambda^t_{U_\alpha }\bigg) + O(t^2)
\end{eqnarray}
Thus, to the first order of evolution time $t$, the entanglement fidelity of the typical DD can be represented by the average entanglement fidelity of the cumulative control unitary operations.

\subsection{Proof of Theorem}

In the scenario where $\varrho_S$ is an $N$-qubit pure state $|\psi\rangle\langle\psi|_S$, the entanglement fidelity of the typical DD after time $t$ can be approximated as
\begin{eqnarray}
    F_e\big(|\psi\rangle\langle\psi|_S, \Lambda^t_\text{BB}\big) \approx \frac{1}{m}\sum_{\alpha=1}^m F_e\big(|\psi\rangle\langle\psi|_S, \Lambda^t_{U_\alpha }\big) + O(t^2). \nonumber
\end{eqnarray}
Note that, for the decoherence channel ${\cal E}^t={\cal E}^t\otimes {\cal I}_{S/i}$, the first and second derivatives $d{\cal E}^t(|\psi\rangle\langle\psi|_S)/dt$ and $d^2{\cal E}^t(|\psi\rangle\langle\psi|_S)/dt^2$ have finite norms. Similarly, the entanglement fidelity of the concatenation of $\Lambda^{\Delta t}_{U_d}$ becomes
\begin{eqnarray}
   &&F_e(|\psi\rangle\langle\psi|_S,\Lambda^{\Delta t}_{U_d,m}\circ\Lambda^{\Delta t}_{U_d,m-1}\circ\Lambda^{\Delta t}_{U_d,1}) \approx \frac{1}{m}\sum_{\alpha=1}^m F_e(|\psi\rangle\langle\psi|, \Lambda^t_{U_d,\alpha}) + O(t^2).
\end{eqnarray}
As $U_dU_d^\dagger = I$, the concatenation of channels $\Lambda^{\Delta t}_{U_d}$ is equivalent to the single operation of a MDD sequence for time $t$, i,e, $\Lambda^{\Delta t}_{U_d,m}\circ\Lambda^{\Delta t}_{U_d,m-1}\circ\cdots\circ\Lambda^{\Delta t}_{U_d,1} = \Lambda^t_\text{MDD}$, and so does the entanglement fidelity;
\begin{eqnarray}
   &&F_e\big(|\psi\rangle\langle\psi|_S, \Lambda^t_\text{MDD}\big)=  F_e\big(|\psi\rangle\langle\psi|_S,\Lambda^{\Delta t}_{U_d,m}\circ\Lambda^{\Delta t}_{U_d,m-1}\circ\cdots\circ\Lambda^{\Delta t}_{U_d,1}\big). \nonumber
\end{eqnarray}

As we showed in Lemma, $\Lambda_\text{MDD}$ can achieve the maximum value of entanglement fidelity over any channel $\Lambda_{U}$ for any time $t$, so the following holds
\begin{eqnarray}
    F_e\big(|\psi\rangle\langle\psi|_S, \Lambda^t_{U_d,\alpha}\big) \ge F_e\big(|\psi\rangle\langle\psi|, \Lambda^t_{U_\alpha }\big),~\forall \alpha.
\end{eqnarray}
This implies that
\begin{eqnarray}
    &&\frac{1}{m}\sum_{\alpha=1}^m F_e\big(|\psi\rangle\langle\psi|, \Lambda^t_{U_d,\alpha}\big) + O(t^2)\ge \frac{1}{m}\sum_{\alpha=1}^m F_e\big(|\psi\rangle\langle\psi|, \Lambda^t_{U_\alpha }\big) + O(t^2).
\end{eqnarray}
Thus, we can conclude that, to the first order of evolution time $t$, the entanglement fidelity of MDD is never less than the typical DD based on the bang-bang operations, i.e., $F_e\big(|\psi\rangle\langle\psi|_S, \Lambda^t_\text{MDD}\big) + O(t^2) \ge F_e\big(|\psi\rangle\langle\psi|_S, \Lambda^t_\text{BB}\big)$.\hfill\qedsymbol

\subsection{Generalization to multiple noisy subsystems}

We here show that MDD is the optimal strategy over the typical DD scheme in an extended scenario where an $N$-qubit pure state has multiple noisy subsystems. Let $k$ subsystems of the $N$-qubit pure state $|\psi\rangle\langle\psi|_S$ be exposed to the decoherence. We assume that the decoherence channles are independent each other, so the $j$-th subsystem evolves according to a channel $\Lambda_{\text{BB},j}$ with $m_j$ bang-bang operations $\{g_{\alpha_j}\}_{\alpha_j=1}^{m_j}$. Then, in the first order of each evolution time $t_j$, the dynamically decoupled state can be approximated as
\begin{eqnarray}
    &&{\cal T}\!\!\prod_{j=1}^k \Lambda_{\text{BB},j}\otimes{\cal I}_{S/j} (|\psi\rangle\langle\psi|_S)\approx \sum_{i=j}^k\left[\frac{1}{m_j}\sum_{\alpha_j=1}^{m_j} \Lambda^{t_j}_{U_{\alpha_j}}(|\psi\rangle\langle\psi |_S) +O( t_j^2)\right],
\end{eqnarray}
where $\Delta t_j=t_j/m_j$.

By the representation in~\eqref{eq:fidelity_approx}, the fidelity between the original state and the dynamically decoupled state, denoted by $F$, is bounded by the fidelity of MDD, $F_\text{MDD}$, from above:
\begin{eqnarray}
    F&=&\langle\psi|{\cal T}\!\!\prod_{j=1}^k \Lambda_{\text{BB},j}\otimes{\cal I}_{S/j} (|\psi\rangle\langle\psi|_S)|\psi\rangle_S \nonumber\\
    &\approx& \sum_{j=1}^k\left[\frac{1}{m_j}\sum_{\alpha_j=1}^{m_j} F_{e}\big(|\psi\rangle\langle\psi |_S,\Lambda^{t_j}_{U_{\alpha_j}}\big) +O(t_j^2)\right]\nonumber\\
    &\le& \sum_{j=1}^k \left[F_{e}\big(|\psi\rangle\langle\psi |_S,\Lambda^{t_j}_{\text{MDD},j}\big) + O(t_j^2) \right] \nonumber\\
    &\approx&  \langle\psi|{\cal T}\!\!\prod_{j=1}^k \Lambda_{\text{MDD},j}\otimes{\cal I}_{S/j} (|\psi\rangle\langle\psi|_S)|\psi\rangle_S + \sum_{j=1}^k O(t^2_j)=F_\text{MDD} + \sum_{j=1}^k O(t^2_j).
\end{eqnarray}
Thus, to the first order of evolution times $t_j$s, the fidelity of MDD is the upper bound of the fidelities which can be obtained by applying the bang-bang operations.

The decoherence channels can be correlated when idle times overlap or the respective subsystems are exposed to a collective decoherence by a common bath and crosstalk. The suppression of correlated decoherence has a significant importance in preserving the performance of quantum algorithms~\cite{Tripathi2022CrossTalk,Coote2025QCTRL}. We will consider a two-qubit MDD scheme for suppressing the crosstalk later.

\begin{figure*}[t!]
    \centering
    \includegraphics[width=\linewidth]{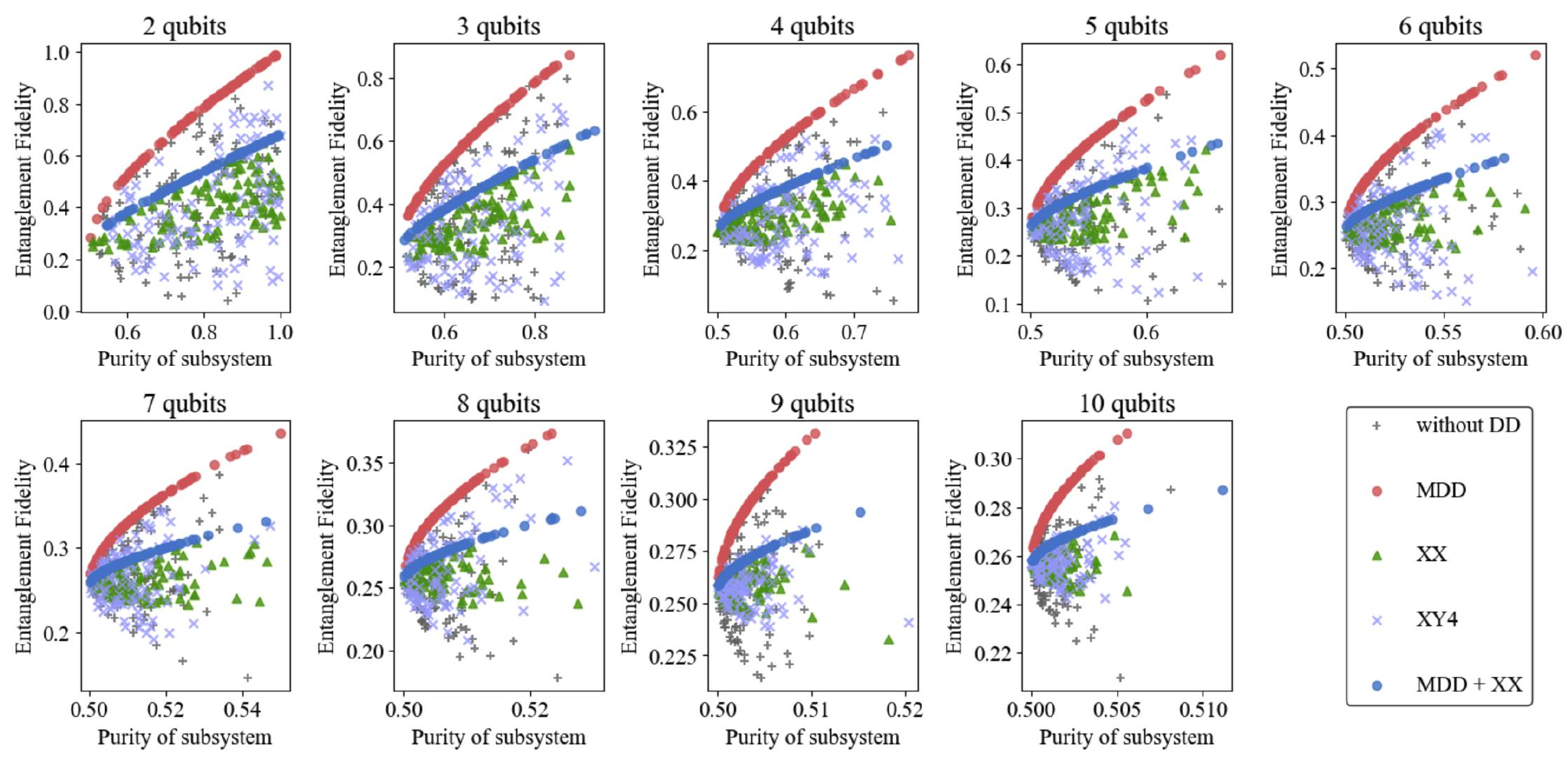}
    \caption{Numerical comparison among various DD schemes: MDD, $XX$, $XY4$, and MDD$+$$XX$. For a randomly sampled hundred states, we assume that one of single-qubit subsystems experience the decoherence given by the relaxation and dephasing channel. We consider $T_1 = 250\mu s$, $T_2=170\mu s$, and the total idle time $t=1000\mu s$. The relaxation and dephasing times are set to the values similar to those of IBM Eagle processor \textit{ibm\_yonsei}.}
    \label{fig:DDcomparison}
\end{figure*}

\begin{figure*}[t!]
    \centering
    \includegraphics[width=\linewidth]{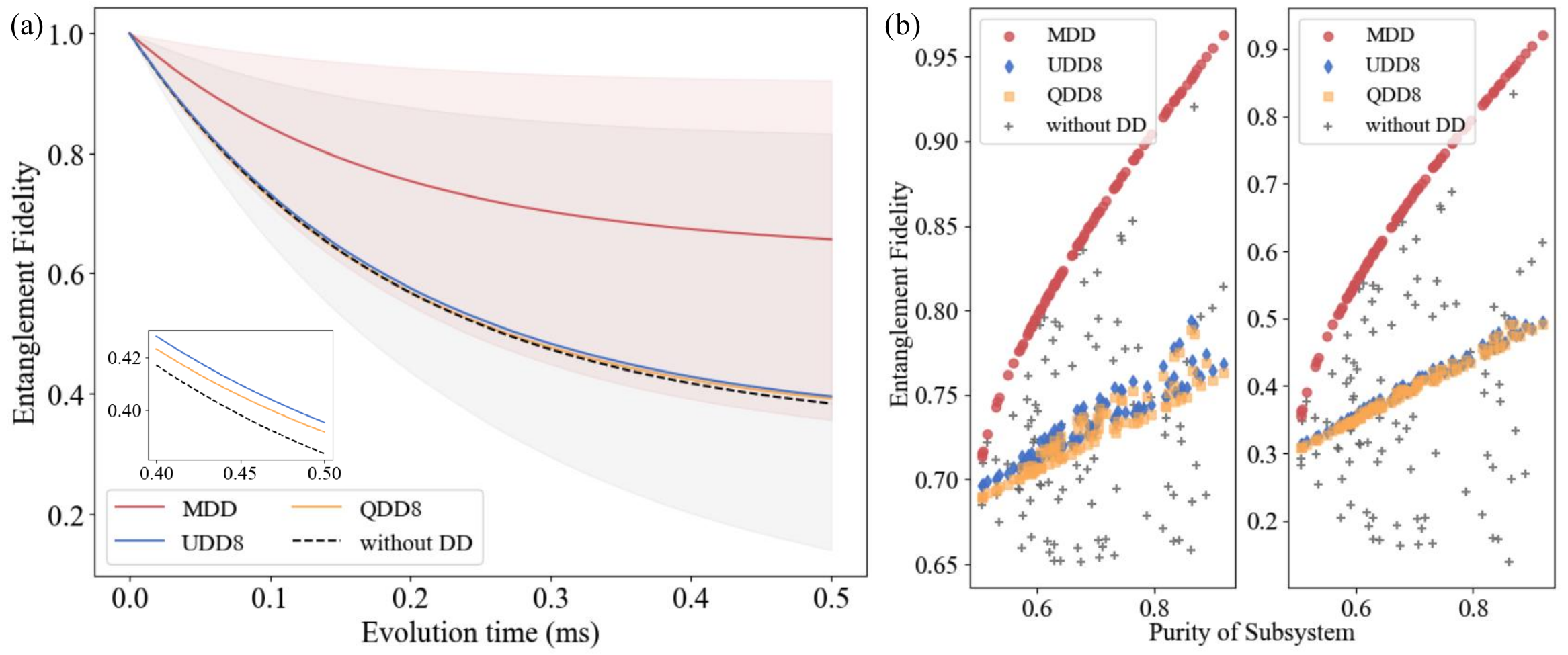}
    \caption{Numerical comparison among MDD, Uhrig DD (UDD), and quadratic DD (QDD). (a) shows the fidelity decay according to the time evolution $t$. We randomly sample a hundred of $3$-qubit states and assume that one of single-qubit subsystems experience the decoherence given by the relaxation and dephasing channel. We consider $T_1 = 250\mu s$, $T_2=170\mu s$. The relaxation and dephasing times are set to the values similar to those of IBM Eagle processor \textit{ibm\_yonsei}. The lines are obtained by averaging the values of entanglement fidelity over the same random states. The red (MDD) and gray (without DD) shaded regions represent the maximum and minimum values of fidelity across these states. (b) shows the distribution of the entanglement fidelity over the purity of subsystems for the same random states used in (a). The left is the results obtained at $t=100 \mu s$ and the right is obtained at $t=500 \mu s$.}
    \label{fig:vsQUDD}
\end{figure*}

\subsection{Numerical comparison}
We compare the performance of the MDD with the typical DDs such as $XX$ and $XY4$. We also consider a new one which inserts $XX$ pulses in between the MDD sequence, called MDD$+$$XX$. We randomly draw a hundred of $N$-qubit pure states and assume that only a single qubit out of $N$ qubits is exposed to the decoherence. The results up to $10$-qubit states are shown in Fig.~\ref{fig:DDcomparison}. Specifically, the sequences of DDs considered are the following~\cite{Ezzell2023}:
\begin{eqnarray}
    \text{MDD}&:&~ U_d - f_t -  U^\dagger_d \\
    \text{$XX$}&:&~ f_{t/4}- X - f_{t/2} -  X - f_{t/4}\nonumber\\
    \text{$XY4$}&:&~ Y - f_{t/4} -  X - f_{t/4} - Y - f_{t/4} - X - f_{t/4}\nonumber\\
    \text{MDD$+XX$}&:&~ U_d - f_{t/4}- X - f_{t/2} -  X - f_{t/4} -  U^\dagger_d, \nonumber
\end{eqnarray}
where $f_\tau$ stands for the free evolution during a time $\tau$. In this simulation, we neglect the time required to implement the gate sequences. In real experiments, we take into account the time consumed by the gate sequences, and the idle times are reduced accordingly. We plot the entanglement fidelity with respect to the purity of the reduced density matrix to highlight that the noisy subsystem is entangled with the remaining parts of the input pure state. The result of MDD$+$$XX$ shows that the intervention of a sub-optimal gate sequence between the MDD sequence decreases the entanglement fidelity. In Fig.~\ref{fig:vsQUDD}, we compare MDD with Uhrig DD (UDD) and quadratic DD (QDD). UDD is proposed to decouple the pure dephasing noise and QDD is a nested UDD scheme which uses two types of pulses to decouple multi-axial relaxation. For a total evolution time $t$ and pulse index $i\in[n]$ with even number $n$, UDD$n$ applies Y-pulses at $t_i = t\sin^2(i\pi/(2n+2))$, and QDD$n$ additionally applies $n$ $X$-pulses at $t_{ij} = t_i + (t_{i+1}-t_i)\sin^2(j\pi/(2n+2))$ between the Y-pulses of the UDD$n$. (See Ref.~\cite{Ezzell2023} for a detailed scheme of UDD and QDD with the performance survey on a superconducting qubit.) For all the cases considered, the MDD outperforms the typical DDs and MDD$+$$XX$.

\begin{figure*}[b!]
    \centering
    \includegraphics[width=\linewidth]{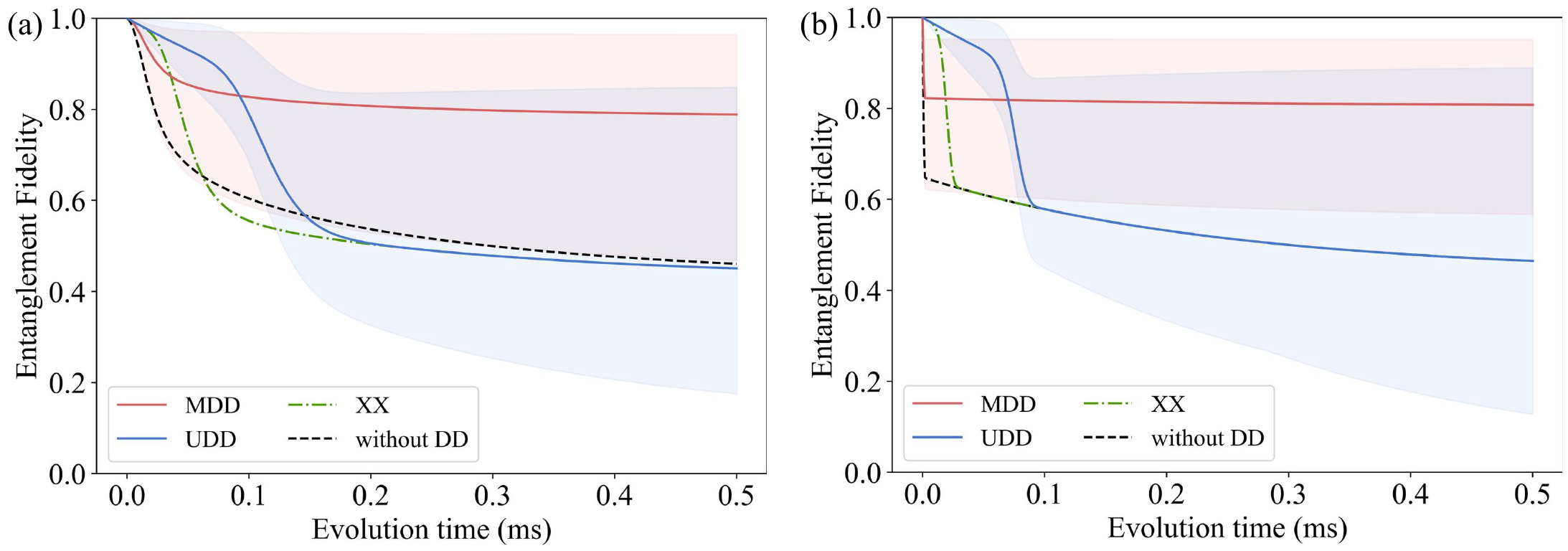}
    \caption{Comparison of MDD, $XX$, UDD, and without DD for two-qubit states ($N=2$) under the amplitude damping ($T_1=250\mu s$) and the dephasing induced by the random phase noise. We consider the Ohmic noise in (a) and the $1/f$ noise in (b) with the cutoff frequency $\omega_c=0.1 (\mu s)^{-1}$, $(\omega_c/2\pi \approx 15.9 \text{kHz})$. The lines are the average values over $20$ randomly sampled states. The red (MDD) and blue (UDD) shaded regions represent the maximum and minimum values of fidelity across these states.}
    \label{fig:random_dephasing}
\end{figure*}

We also consider the random dephasing noise induced by the Hamiltonian, $\frac{1}{2}\beta(t)\sigma_{z,i}$, acting on the $i$-th subsystem of a two-qubit state, together with the amplitude damping. Assuming that the average of the accumulated random phase is zero, the phase decays with $\exp(-\chi(t))$, where $\chi(t)$ is determined by the spectral density of the random noise $S(\omega)$ and the filer function $F(\omega t)$ for the angular frequency $\omega$~\cite{Uys2009,Ezzell2023} as
\begin{eqnarray}
    \chi(t) = \frac{2}{\pi}\int_0^\infty \frac{S(\omega)}{\omega}F(\omega t) d\omega.
\end{eqnarray}
The operation of DD can be described by the sign change of the filter function: For n $\pi$-pulses acting at times $t_j$, $F(\omega t)$ becomes
\begin{eqnarray}
    F(\omega t) = \left\vert 1 + (-1)^{n+1} +2\sum_{j=1}^n (-1)^j e^{i\omega t_j}\right\vert^2.
\end{eqnarray}
We consider two noise models: (i) Ohmic spectral density $S_O(\omega)=\omega e^{-(\omega/\omega_c)^2}$~\cite{Leggett1987Ohm}, and (ii) $1/f$ spectral density $S_{1/f}(\omega)= \omega^{-1}e^{-(\omega/\omega_c)^2}$~\cite{Yoshihara2006,Paladino2014noise1f}, where $\omega_\text{c}$ is the cutoff frequency. In the Ohmic noise, high-frequency region is dominated, while in the $1/f$ noise, low-frequency region is dominated. Figure~\ref{fig:random_dephasing} shows the comparisons among MDD, $XX$, UDD$8$, and without DD. This result demonstrates that MDD can preserve the fidelity for a longer duration compared to UDD and $XX$ sequences.

\section{The minimum decay rate of the entanglement fidelity}

A well-designed DD scheme should have a low fidelity decay rate to preserve the initial quantum information for a long time. We here show that, for the uncorrelated decoherence, the MDD sequence is the optimal strategy to minimize the decay rate of the entanglement fidelity among all sequences of single-qubit unitary operations $(U,U^\dagger)$. We also consider a correlated decoherence model and extend the MDD scheme to applying $2$-qubit gate sequence to suppress crosstalk.

\subsection{Uncorrelated decoherence}

We consider the same scenario as considered in the theorem; a $N$-qubit pure state $|\psi\rangle\langle\psi|_S$ and the $i$-th subsystem is exposed to the decoherence. In this scenario, the decay rate of the entanglement fidelity for a pair of single-qubit unitary operations $(U_i,U_i^\dagger)$ acting on the subsystem is defined by
\begin{eqnarray}
    \frac{dF_{e,U}}{dt} &=& \frac{d}{dt}\langle\psi| \Lambda_U^t (|\psi\rangle\langle\psi|_S)|\psi\rangle_S \nonumber\\
    &=&\frac{d}{dt}\langle\psi| U_i^\dagger{\cal E}^t (U_i|\psi\rangle\langle\psi|_S U_i^\dagger)U_i|\psi\rangle_S \nonumber\\
    &=&\langle\psi| U_i^\dagger{\cal L} (U_i|\psi\rangle\langle\psi|_S U_i^\dagger)U_i|\psi\rangle_S 
\end{eqnarray}
where ${\cal L}$ is the superoperator, which governs the time evolution of the state with the master equation:
\begin{eqnarray}
    \frac{d}{dt}{U|\psi\rangle\langle\psi|_SU^\dagger} 
    &=& -i\big[H_0,U|\psi\rangle\langle\psi|_SU^\dagger\big] + \sum_k \Gamma_k\left( L_k U|\psi\rangle\langle\psi|_SU^\dagger L^\dagger_k - \frac{1}{2}\left\{L^\dagger_kL_k,U|\psi\rangle\langle\psi|_SU^\dagger\right\}\right).\nonumber \\
    &=& {\cal L}(U|\psi\rangle\langle\psi|_S U^\dagger).
\end{eqnarray}

For the combined decoherence of the amplitude damping (AD) and dephasing (PD), the jump operators $\{L_i\}$ become
\begin{eqnarray}
    L_1^\text{AD} = \frac{1}{2}(\sigma_{x,i}+i\sigma_{y,i})\otimes I_{S/i}\quad\text{and}\quad L_2^\text{PD} = \sigma_{z,i}\otimes I_{S/i},\nonumber
\end{eqnarray}
for $\Gamma_1 = 1/T_1$ and $\Gamma_2 = 1/T_p$. The variance of each jump operator $L$ determines the first derivative of the entanglement fidelity as
\begin{eqnarray}
    \frac{dF_{e,U}}{dt} &=& -\sum_{k=1}^2 \Gamma_k \left(\langle\psi|U^\dagger L^\dagger_k L_k U|\psi\rangle - \vert \langle\psi|U^\dagger L_k U|\psi\rangle \vert^2\right) \nonumber\\
    &=& -\sum_{k=1}^2 \Gamma_k\text{Var}_{\sigma_i}[UL_kU^\dagger],
\end{eqnarray}
where $\sigma_i$ is the density matrix of the $i$-th subsystem. As the variance is positive semidefinite, the fidelity gradually decreases by the time evolution.

The fidelity decay rate is then given by a quadratic function with the $z$-component of the Bloch vector of the reduced density matrix $\sigma_U = U\sigma_iU^\dagger$, $r_{z,U}$:
\begin{eqnarray}
    \left\vert\frac{dF_{e,U}}{dt}\right\vert &=& \sum_{k=1}^2 \Gamma_k\text{Var}_{\sigma_i}[UL_kU^\dagger] \\
    &=&\sum_{k=1}^2 \Gamma_k \left[\Tr(\sigma_U L^\dagger_{k,i} L_{k,i})  - \vert \Tr\left(\sigma_U L_{k,i} \right) \vert^2\right] \nonumber\\
    &=& \Gamma_1\left[\frac{1}{2}(1-r_{z,U})-\frac{1}{4}(r^2-r_{z,U}^2)\right] + \Gamma_2\left(1-r_{z,U}^2\right). \nonumber
\end{eqnarray}
Note that this function is positive semidefinite in the entire domain $r_{z,U}\in[-r,r]$. When $d^2\left\vert\frac{dF}{dt}\right\vert/dr_{z,U}^2=(\Gamma_1-4\Gamma_2)/2 > 0 $, the extreme point becomes $r_{z,U}^\star = \Gamma_1/(\Gamma_1-4\Gamma_2) >1$, so the minimum decay rate is obtained at $r_{z,U}=r$. When $d^2\left\vert\frac{dF}{dt}\right\vert/dr_{z,U}^2 < 0 $, the extreme point $r_{z,U}^\star$ appears in the negative values, so the minimum decay rate is obtained at $r_{z,U}=r$. When $d^2\left\vert\frac{dF}{dt}\right\vert/dr_{z,U}^2=0$, the decay rate becomes a linear function that is monotonically decreasing in the domain, so the minimum value is obtained at $r_{z,U}=r$. In other words,
\begin{eqnarray}
     U_d = \text{argmin}_U \left\vert\frac{dF_{e,U}}{dt}\right\vert,
\end{eqnarray}
where $U_d$ is the unitary operator which aligns the Bloch vector of the $i$-th reduced density matrix in $r_{z,U}=r$ and its eigenvalues in descending order. This implies that applying $U_d$ can delay the passage time in which a state evolves to the ground state. We here investigate the upper bound of the evolution time achievable via a pair of single-qubit unitary operations under the specific decoherence model.

\subsection{Correlated decoherence}

Consider that the $i$-th and $j$-th subsystems represented by a density matrix $\sigma_{ij}$ are exposed to the decoherence caused by the amplitude damping, dephasing, and $ZZ$ crosstalk. The jump operator of the $ZZ$ crosstalk is given by
\begin{eqnarray}
    L_{zz} = \sigma_{z,i} \otimes \sigma_{z,j} \otimes I_{S/ij}.
\end{eqnarray}
The fidelity decay rate becomes
\begin{eqnarray}
    \left\vert\frac{dF_e}{dt}\right\vert &=& \sum_{k=1}^2 \sum_{l\in\{i,j\}}\Gamma_{k,l}\text{Var}_{\sigma_{ij}}[L_{k,l}] + \Gamma_{zz}\text{Var}_{\sigma_{ij}}[L_{zz}].
\end{eqnarray}
In this case, MDD can generalize to applying a pair of $2$-qubit unitary operations $(U_{ij},U_{ij}^\dagger)$ to transform the density matrix of the noisy subsystems to a target (ansatz) state defined by
\begin{eqnarray}
    \tilde{\sigma}_{ij} &=& U_{ij}\sigma_{ij}U_{ij}^\dagger \nonumber\\
    &=&\frac{1}{4}\left(I\otimes I + c_{1}\sigma_z\otimes I + c_2I\otimes\sigma_z + c_3\sigma_z\otimes\sigma_z\right).
\end{eqnarray}
Note that this state is positive when $1+c_1+c_2+c_3>0$, $1+c_1-c_2-c_3>0$, $1-c_1+c_2-c_3>0$, and $1-c_1-c_2+c_3>0$. For this state, the decay rate can be read as
\begin{eqnarray}
    \left\vert\frac{dF}{dt}\right\vert = \Gamma_{1,i}\left[\frac{1}{2}(1-c_1)-\frac{1}{4}(r_i^2-c_1^2)\right] + \Gamma_{2,i}\left(1-c_1^2\right) + \Gamma_{1,j}\left[\frac{1}{2}(1-c_2)-\frac{1}{4}(r_j^2-c_2^2)\right] + \Gamma_{2,j}\left(1-c_2^2\right) + \Gamma_{zz}(1-c_3^2),
\end{eqnarray}
where $r_x$ is the norm and the Bloch vector of the $x$-th subsystem.

The MDD sequence of the minimum fidelity decay rate can be obtained by solving following optimization,
\begin{eqnarray}
    \min_{c_1,c_2,c_3} \left\vert\frac{dF}{dt}\right\vert~~\text{subject to}~~\tilde{\sigma}_{ij}>0,
\end{eqnarray}
and deducing the gate implementation of the unitary transformation from $\sigma_{ij}$ to $\tilde{\sigma}_{ij}$ with the optimal coefficients. In particular, for a pure state $\sigma_{ij}$, the target state is given by $|00\rangle$, where the optimal coefficients are given by $c_1=1$, $c_2=1$, and $c_3=1$, and the fidelity decay rate becomes zero. For a general mixed state, MDD cannot completely suppresses the crosstalk, as putting $c_3=1$ to eliminate the crosstalk effect leads to contradictory conditions, $c_1>c_2$ and $c_2>c_1$, from the positivity of target state.

The implementation of the two-qubit MDD sequence generally requires entangling gates. However, finding the $2$-qubit MDD sequence within a given idle time is a nontrivial problem, and such $2$-qubit gates may introduce additional idle qubits elsewhere in the quantum circuit. Thus, in a weak-crosstalk regime, applying the single-qubit MDD sequence at each idle qubit can be an efficient heuristic approach.

\section{Experimental details}

\subsection{Quantum Fourier Transform}

\begin{figure*}[b!] 
    \centering
    \includegraphics[width=0.9\linewidth]{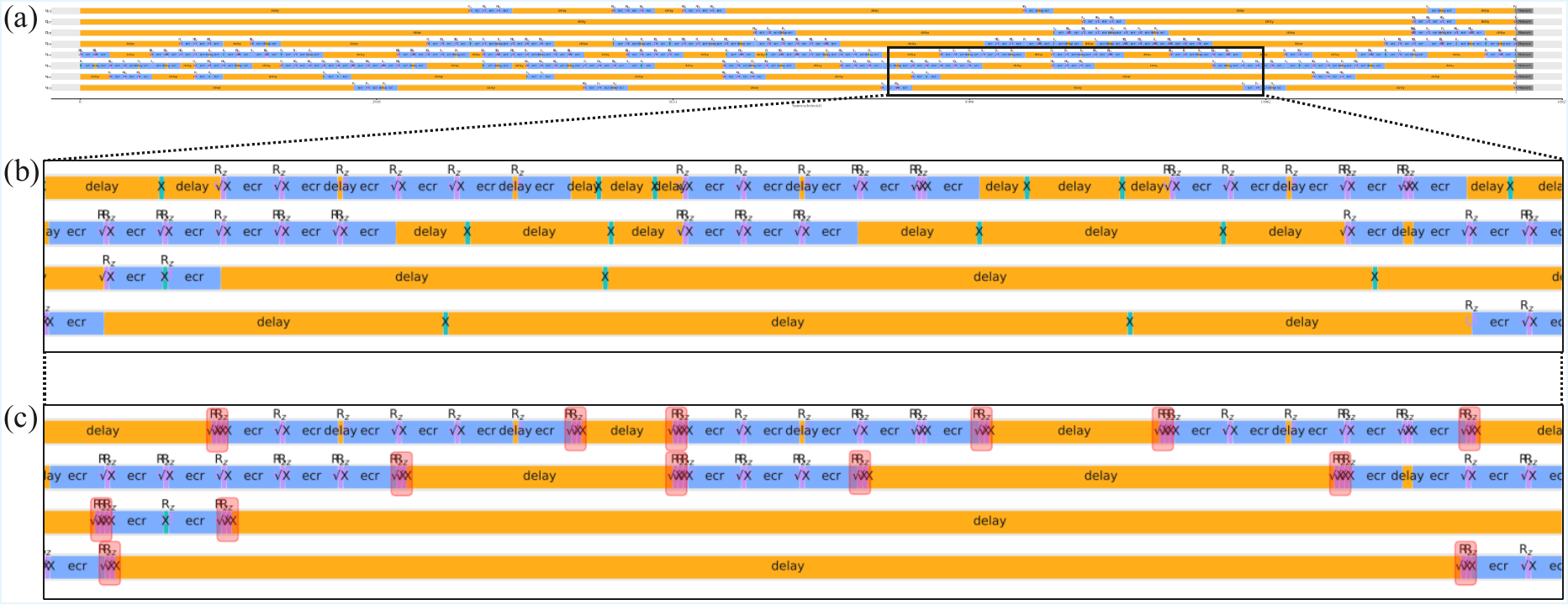}
    \caption{
    Transpiled circuit schedule for an eight-qubit QFT. (a) The overall circuit timeline without DD. (b) Representative region of the circuit with a $XX$, where the $X$ gates are shown as green blocks. (c) The same region with MDD, where the MDD gates are shown in magenta.
    }
    \label{fig:fig_t_qft}
\end{figure*}

A key challenge in implementing the Quantum Fourier Transform (QFT) on superconducting quantum hardware arises from the inherent mismatch between the algorithm’s logical connectivity and the device’s physical topology. The QFT requires all-to-all qubit interactions through controlled-phase operations, whereas the IBM Eagle processor features a heavy-hexagonal lattice topology that limits each qubit to nearest-neighbor couplings. To address this mismatch, the Qiskit transpiler introduces SWAP operations that route quantum states across the lattice to enable temporary adjacency between distant qubits. Each SWAP decomposes into three CNOT gates, substantially increasing the total two-qubit gate count and circuit depth. The resulting deeper circuits contain more frequent and longer idle periods, during which qubits accumulate dephasing and crosstalk errors. Consequently, the non-local structure of QFT and the restricted qubit-connectivity amplify both coherent and incoherent error channels, posing a major scalability bottleneck for fault-prone processors.

The mapping from the QFT circuits to the physical layout on the devices was performed using the Qiskit's circuit transpiler with optimization levels $3$. For temporal scheduling, we employed the default as-late-as-possible (ALAP) scheduling method~\cite{javadi2024quantum}. This strategy is a key component of the Qiskit scheduling model, as it inherently mitigates $T_1$ and $T_2$ decay errors. It achieves this by maximizing the time qubits spend in their initial ground-state before the first operation and minimizing the idle time between the final gate and the measurement.

Figure~\ref{fig:fig_t_qft} shows the transpiled circuit timeline for an eight-qubit QFT experiment performed on the \textit{ibm\_yonsei} device, illustrating the structure under three conditions: (i) without DD, (ii) $XX$, and (iii) MDD. While panel (a) displays the complete timeline of the baseline circuit, panels (b) and (c) highlight representative segments where $XX$ and MDD were applied.

\subsection{Sample-based Quantum Diagonalization}

\begin{figure*}[t] 
    \centering
    \includegraphics[width=.9\linewidth]{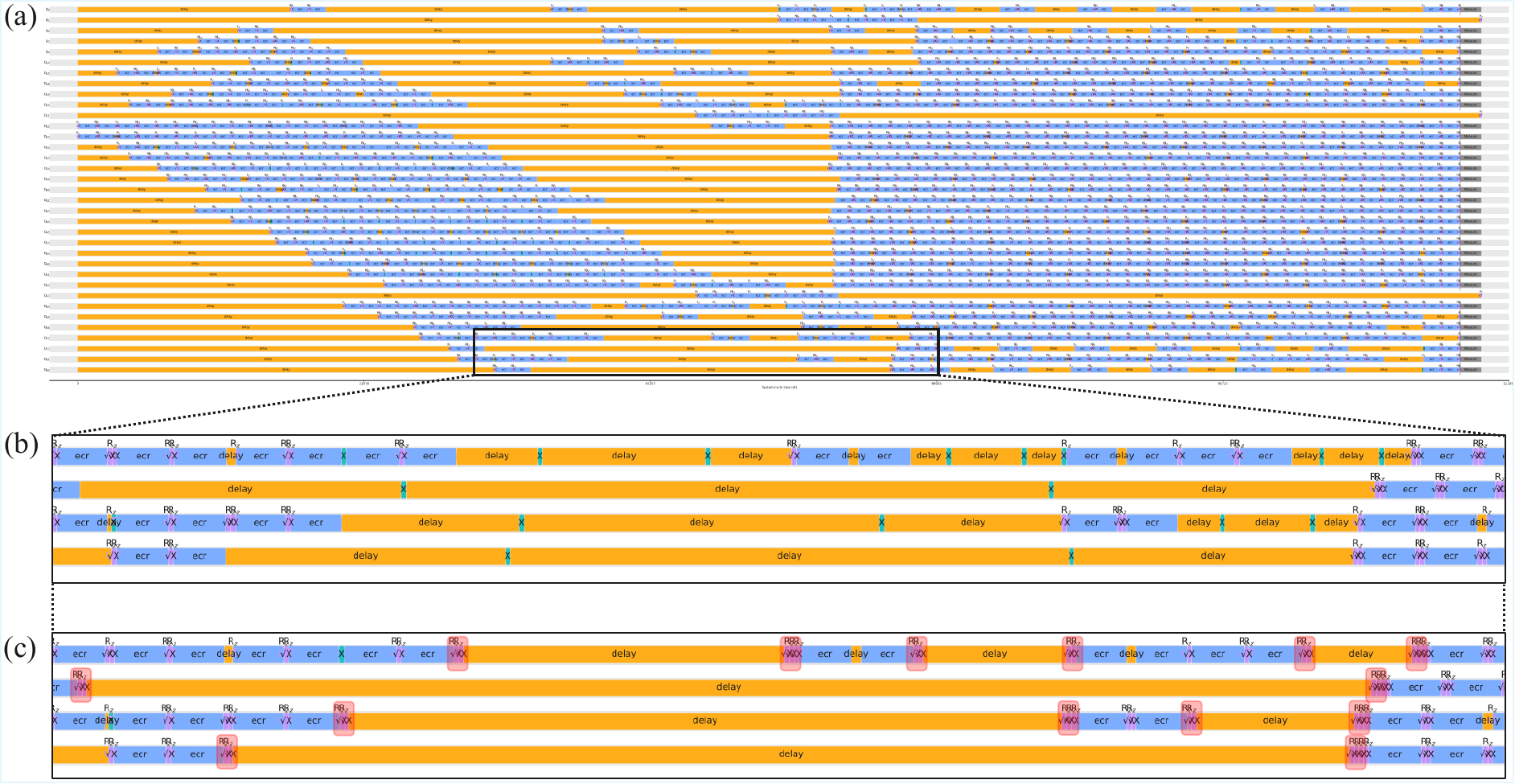}
    \caption{
    Transpiled circuit schedule for an $35$-qubit SQD. (a) The overall circuit timeline without DD. (b) Representative region of the circuit with a $XX$, where the $X$ gates are shown as green blocks. (c) The same region with MDD, where the MDD gates are shown in magenta.
    }
    \label{fig:fig_t_sqd}
\end{figure*}

The Sample-based Quantum Diagonalization (SQD) algorithm combines quantum state sampling with classical subspace diagonalization to estimate the eigenenergies and eigenstates of molecular Hamiltonians~\cite{Javier2025SQD}. This hybrid quantum–classical approach is robust to hardware noise and scalable to large systems with millions of Hamiltonian terms. By leveraging efficiently prepared quantum samples and classical post-processing, SQD enables accurate ground-state energy estimation on current noisy intermediate-scale quantum (NISQ) devices~\cite{preskill2018quantum}. The SQD experiments described in the main text estimated the ground-state energy of the nitrogen molecule ($\mathrm{N}_2$) using $35$- and $56$-qubit circuits derived from the $6–31$G and cc-pVDZ basis sets, respectively. All experiments were executed on \textit{ibm\_yonsei} processor, with circuits transpiled using Qiskit’s circuit transpiler with optimization levels $3$ and scheduled using the ALAP.

\textbf{Molecular Hamiltonian and Encoding.---}The molecular Hamiltonian is expressed in the second-quantized form as
\begin{eqnarray}
    \hat{H} =
    \sum_{p r, \sigma} h_{p r} \, \hat{a}^{\dagger}_{p \sigma} \hat{a}_{r \sigma}
    + \frac{1}{2} \sum_{p r q s, \sigma \tau} (p r | q s) \,
    \hat{a}^{\dagger}_{p \sigma} \hat{a}^{\dagger}_{q \tau}
    \hat{a}_{s \tau} \hat{a}_{r \sigma},
    \label{eq:hamiltonian}
\end{eqnarray}
where $\hat{a}^{\dagger}_{p \sigma}$ ($\hat{a}_{p \sigma}$) denotes the fermionic creation (annihilation) operator for the $p$-th spatial orbital with spin $\sigma$. The coefficients $h_{p r}$ and $(p r | q s)$ represent the one- and two-electron integrals, respectively, which were computed using the \textsc{PySCF} package~\cite{sun2018pyscf}. Two basis sets were considered: (i) $6-31$G, consisting of $16$ spatial orbitals and (ii) cc-pVDZ, consisting of $26$ spatial orbitals. Under the Jordan–Wigner (JW) mapping~\cite{jordan1928paulische}, these systems map to $32$ and $52$ spin orbitals, respectively. Including ancillary qubits, the resulting circuit sizes were $35$-qubits ($32$ system qubits $+$ $3$ ancillary qubits) for the $6–31$G basis and $56$-qubits ($52$ system qubits $+$ $4$ ancillary qubits) for the cc-pVDZ basis. In both experiments, ten valence electrons were distributed equally between spin-up and spin-down orbitals ($N_\uparrow=N_\downarrow=5$), ensuring a spin-singlet state with $S_z = 0$.

\textbf{Quantum Circuit Ansatz.---}The quantum circuits used for sample generation were constructed using the Local Unitary Coupled Cluster Jastrow (LUCJ) Ansatz~\cite{motta2023bridging}. LUCJ is a hardware-efficient variant of the Unitary Coupled Cluster Jastrow (UCJ) Ansatz, designed for the limited qubit connectivity of superconducting processors. It preserves the physical interpretability of the UCJ form while improving hardware efficiency through local gate construction.

The LUCJ Ansatz consists of $L$ layers, each composed of an orbital-rotation block ($e^{K_\mu}$, $e^{-K_\mu}$) surrounding a diagonal Coulomb interaction ($e^{iJ_\mu}$):
\begin{eqnarray}
    |\psi\rangle = \prod_{\mu=1}^{L} \big( e^{K_\mu} e^{iJ_\mu} e^{-K_\mu} \big) |\Phi_0\rangle,
\end{eqnarray}
where $|\Phi_0\rangle$ is the reference Hartree–Fock state.  

The orbital-rotation unitaries $e^{\pm K_\mu}$ are implemented through givens rotation gates acting on adjacent spin orbitals, decomposed into $(XX + YY)$ entangling gates and single-qubit $R_Z$ rotations. This structure aligns naturally with heavy-hexagonal topology of IBM devices and eliminates the need for additional SWAP operations. The diagonal Coulomb term $e^{iJ_\mu}$ represents number–number ($U_{nn}$) interactions. In the original UCJ formulation, $J_{\alpha\alpha}$ and $J_{\alpha\beta}$ matrices couple all orbital pairs, which would require long-range $U_{nn}$ gates and numerous SWAPs. LUCJ introduces sparsity constraints on $J_{\alpha\alpha}$ and $J_{\alpha\beta}$, eliminating interactions incompatible with the device connectivity and thereby reducing circuit depth and gate count.

To further enhance hardware efficiency, the orbital-to-qubit mapping follows a “zig-zag” pattern optimized for heavy-hexagonal topology. Same-spin orbitals ($\alpha-\alpha$ or $\beta-\beta$) are arranged linearly to preserve nearest-neighbor connectivity, while cross-spin ($\alpha-\beta$) interactions are routed through every fourth orbital index (e.g., $0, 4, 8, \dots$), with ancillary qubits used to maintain logical adjacency when needed. All circuit parameters were initialized from classical coupled-cluster singles and doubles (CCSD)~\cite{sun2018pyscf} amplitudes $(t_1, t_2)$, providing physically meaningful starting points that enable high-fidelity state preparation without extensive quantum variational optimization.

\textbf{Compilation.---}To further optimize circuit depth and gate count, we applied the pre-initialization optimization stage provided by the \textit{ffsim} library~\cite{ffsim}. This pass decomposes and merges orbital-rotation blocks before full transpilation, effectively removing redundant single-qubit rotations and simplifying adjacent gate sequences. As a result, the total number of single- and two-qubit gates was reduced in both experiments.

For the $35$-qubit experiment, gate counts were reduced from $4,386$ to $2,413$ for $R_Z$ gates, from $3,354$ to $2,111$ for $SX$ gates, and from $1,366$ to $730$ for ECR gates. For the $56$-qubit experiment, the counts decreased from $14,069$ to $6,297$ for $R_Z$ gates, from $9,954$ to $5,357$ for $SX$ gates, and from $3,798$ to $1,740$ for ECR gates. Overall, the pre-initialization optimization achieved a $45$--$55$\% reduction in single- and two-qubit gate counts across both experiment, substantially lowering the cumulative error rate and execution time. The final compiled circuits had depths of $148$ ($35$-qubit) and $223$ ($56$-qubit).

Fig.~\ref{fig:fig_t_sqd} shows the transpiled circuit timeline for the $35$-qubit SQD experiment performed on the \textit{ibm\_yonsei} device, illustrating the structure under three conditions: (i) without DD, (ii) $XX$, and (iii) MDD. While panel (a) displays the complete timeline of the baseline circuit, panels (b) and (c) highlight representative segments where $XX$ and MDD were applied.

\textbf{Sample-based Diagonalization Procedure.---}The final stage of the SQD workflow performs classical post-processing of noisy bitstring samples $\tilde{\chi}$ obtained from the quantum hardware. Since hardware noise can violate physical symmetries such as fixed particle number and spin-$Z$ conservation, these samples must be corrected before subspace construction. The self-consistent configuration recovery procedure was employed to restore corrupted configurations and enhance the effective signal-to-noise ratio~\cite{Javier2025SQD}.

For each measured bitstring $x$, SQD leverages the fixed particle-number symmetry of the molecular Hamiltonian. The number of occupied orbitals (i.e., the number of $1$'s in $x$) defines the total electron count $N_x$. If $N_x$ differs from the expected particle number $N$, the bitstring is regarded as corrupted by hardware noise. Instead of discarding these invalid configurations through post-selection, the self-consistent configuration recovery procedure probabilistically corrects them using the average orbital occupancy $\mathbf{n}$, whose elements represent the estimated mean occupation probabilities of each spin orbital. When $N_x < N$, a subset of unoccupied orbitals ($0$’s) are flipped to $1$’s to restore the missing electrons; conversely, when $N_x > N$, occupied orbitals ($1$’s) are flipped to $0$’s. The bit-flip probability for orbital $i$ was determined by the deviation $|x_i - n_i|$ between the measured bit value and the mean orbital occupancy, further weighted by a modified ReLU function $w(y)$ to prevent excessive flipping near equilibrium occupations. The weighting function is defined as
\begin{eqnarray}
    w(y) =  \begin{cases}
                \displaystyle \frac{\delta}{h} y, & y \le h, \\
                \displaystyle \delta + (1-\delta)\frac{y-h}{1-h}, & y > h,
            \end{cases}
\end{eqnarray}
where $h = N/M$ represents the filling factor, $M$ is the total number of spin orbitals, and $\delta = 0.01$ sets the smoothness at the transition point.

Since $\mathbf{n}$ is initially unknown, SQD iteratively refines it through a self-consistent feedback loop:
\begin{enumerate}[label=(\arabic*), leftmargin=4em]
    \item Initially, only symmetry-preserving configurations $\tilde{\chi}_{\text{correct}}$ ($N_x = N$) are selected from $\tilde{\chi}$ and divided into $K$ random batches ${S^{(1)}, \dots, S^{(K)}}$, each defining a classical subspace for diagonalization.

    \item For each batch $S^{(k)}$, the molecular Hamiltonian $\hat{H}$ is projected onto the corresponding subspace, $\hat{H}_{S^{(k)}} = P_{S^{(k)}} \hat{H} P_{S^{(k)}}$, and classically diagonalized, yielding approximate eigenstates $\{|\psi^{(k)}\rangle\}$ and eigenvalues $\{E^{(k)}\}$.  

    \item The $\mathbf{n}$ are computed from the ensemble of low-energy eigenstates, forming the first estimate of $\mathbf{n}^{(1)}$. 

    \item In subsequent iterations, $\mathbf{n}^{(t)}$ is used to correct previously invalid configurations $\tilde{\chi}_{\text{incorrect}}$, producing new corrected configurations $\tilde{\chi}_{\text{correct, new}}$. 

    \item The newly recovered configurations, denoted as $\tilde{\chi}_{\text{correct, new}}$, are merged with the previously valid configurations $\tilde{\chi}_{\mathrm{correct}}$ to form the expanded configuration pool, $\tilde{\chi}_{R} = \tilde{\chi}_{\text{correct}} \cup \tilde{\chi}_{\text{correct,new}}$. In the subsequent iteration, new batches are sampled from this expanded pool, followed by Hamiltonian projection and diagonalization for each batch.

    \item The $\mathbf{n}$ is then updated by averaging the mean orbital occupancies computed from all batches, rather than directly averaging the lowest-energy eigenstates. This updated $\mathbf{n}$ serves as a refined estimate for the next iteration. The iterative process continues until a predefined stopping criterion is met, typically based on the convergence of the estimated ground-state energy or the stability of the mean occupations.
    
\end{enumerate}

For both basis experiment of the $\mathrm{N}_2$ molecule, the self-consistent configuration recovery iterations yielded a monotonic convergence of the estimated ground-state energy across iterations, indicating effective noise suppression and consistent subspace refinement. The SQD parameters were as follows: five self-consistent configuration recovery iterations, $K=10$ batches, and $300$ samples per batch for the $35$-qubit experiment, and ten self-consistent configuration recovery iterations with the same batch parameters for the $56$-qubit experiment.

\begin{figure}[h]   
    \centering
    \includegraphics[width=0.8\linewidth]{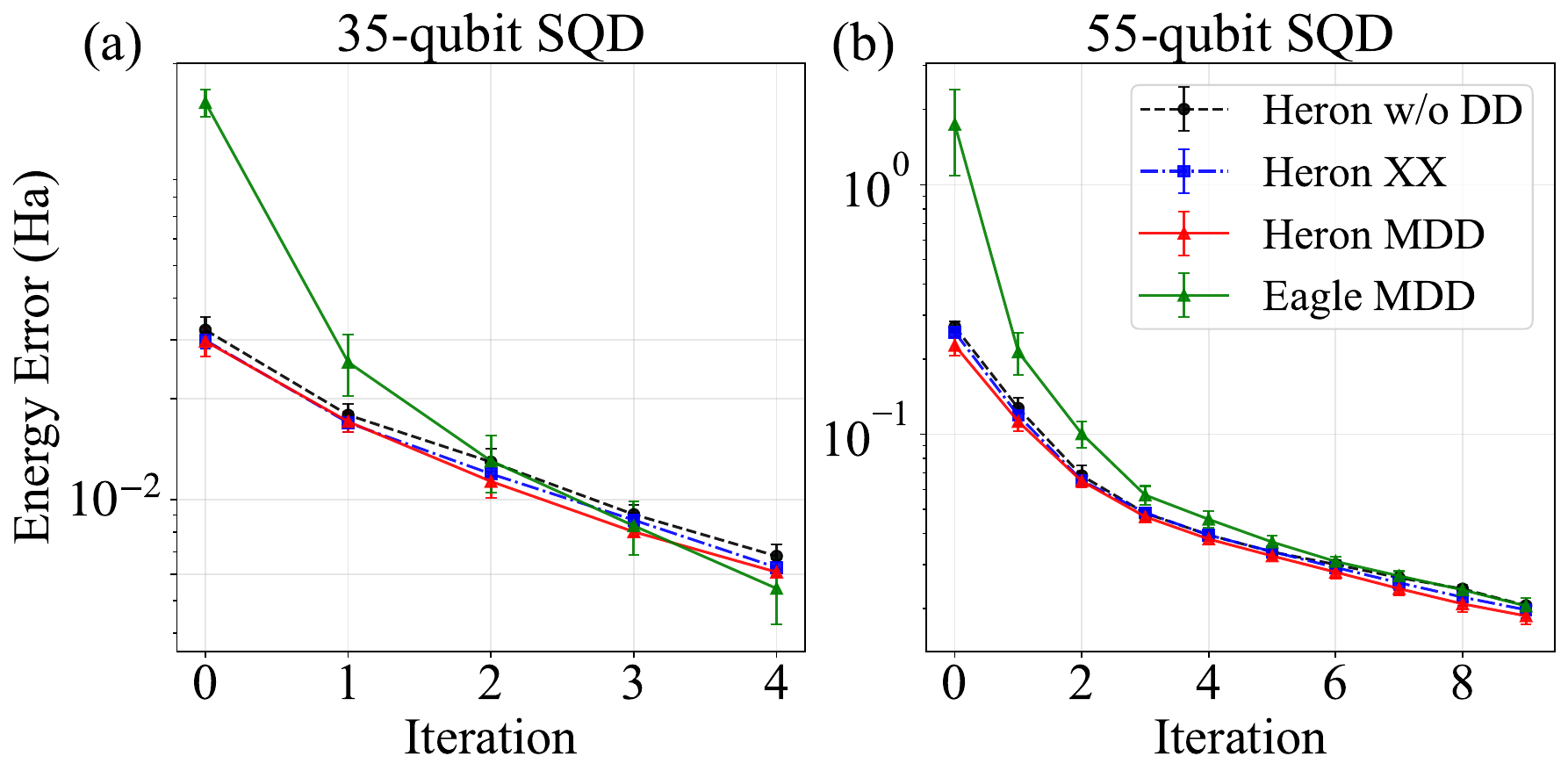}
    \caption{ 
    Comparison of the estimated ground-state energy errors for $\mathrm{N}_2$ obtained using the SQD workflow with three DD schemes: without DD (black dashed line), $XX$ (blue dotdashed line), and MDD (red solid line). Panels (a) and (b) show the mean energy errors versus the number of self-consistent configuration recovery iterations for $35$- and $56$-qubit experiments on the \textit{ibm\_fez} (heron) device, respectively. The corresponding reference results with MDD on the \textit{ibm\_yonsei} (Eagle) processor are shown in green solid line for comparison. Error bars represent the standard deviation over five independent runs.
    }
    \label{fig:fig_heron_sqd_iter}
\end{figure}

\textbf{Experimental Results of Heron processor.---}We conducted additional SQD experiments on IBM $156$-qubit Heron processor \textit{ibm\_fez}. For the cc-pVDZ basis, only three ancillary qubits were required, resulting in a total of $55$ qubits ($52$ system qubits $+$ $3$ ancillary qubits). Both the $35$- and $55$-qubit LUCJ circuits were executed with $10^{5}$ measurement shots, $K = 10$ batches, and $300$ samples per batch, over five and ten self-consistent configuration recovery iterations, respectively.

The Fig.~\ref{fig:fig_heron_sqd_iter} show the iterative convergence of the estimated ground-state energy errors for the $35$- and $55$-qubit experiments conducted on the \textit{ibm\_fez} processor. Consistent with the observations from the \textit{ibm\_yonsei} device, the MDD achieves lower initial energy errors compared to both the without DD and the $XX$. In the $35$-qubit experiment, MDD reduces the initial energy error to $2.97 \times 10^{-2} \pm 2.9 \times 10^{-3}$ Ha, which is approximately $0.64\%$ lower than the $XX$ and $7.40\%$ lower than the without DD. After five self-consistent configuration recovery iterations, MDD further improves the convergence, reaching a final energy error of $6.08 \times 10^{-3} \pm 1.7 \times 10^{-4}$ Ha. This corresponds to an approximately $1.03$ times improvement in accuracy relative to $XX$ and approximately $1.12$ times relative to the without DD.

A similar enhancement is observed for the $55$-qubit experiment. At the initial iteration, MDD yields an energy error of $2.27 \times 10^{-1} \pm 2.02 \times 10^{-2}$ Ha, approximately $11\%$ and $16\%$ lower than the $XX$ and the without DD. As the self-consistent configuration recovery iterations proceed, all methods show a steady decrease in error, but MDD consistently maintains the lowest values. After ten iterations, the final energy error obtained with MDD converges to $1.87 \times 10^{-2} \pm 1.45 \times 10^{-3}$ Ha, representing a approximately $1.06$ times improvement over $XX$ and approximately $1.1$ times relative to the without DD.

\textbf{Comparison between Eagle and Heron processors.---}We compared the performance of MDD on the \textit{ibm\_yonsei} and \textit{ibm\_fez} processors. As shown in Fig.~\ref{fig:fig_heron_sqd_iter}, the MDD results on \textit{ibm\_yonsei} (green) closely follow those on \textit{ibm\_fez} (red). In the $35$-qubit experiment, Heron initially yielded a smaller energy error than Eagle, yet both converged to nearly identical final accuracies of approximately $6\times10^{-3}$ Ha after five self-consistent recovery iterations. A similar trend was observed in the $55$-qubit experiment: although Heron started with an energy error about $7.7$ times lower than Eagle, both ultimately reached comparable final values around $2\times10^{-2}$ Ha after ten iterations. These results demonstrate that MDD effectively suppresses decoherence, enabling the Eagle processor to achieve nearly the same accuracy as the more advanced Heron processor.

\subsection{Idle time analysis}

Idle time is defined as an interval during circuit execution in which a physical qubit is not subject to any active gate and is therefore left idle; during these intervals amplitude damping and dephasing gradually degrade its quantum state. The MDD protocol was applied only to idle intervals exceeding device-specific thresholds—$240$\,ns for \textit{ibm\_yonsei} and $96$\,ns for \textit{ibm\_fez}—chosen empirically based on the native gate durations of each processor.

\begin{figure}[h] 
    \centering
    \includegraphics[width=0.7\linewidth]{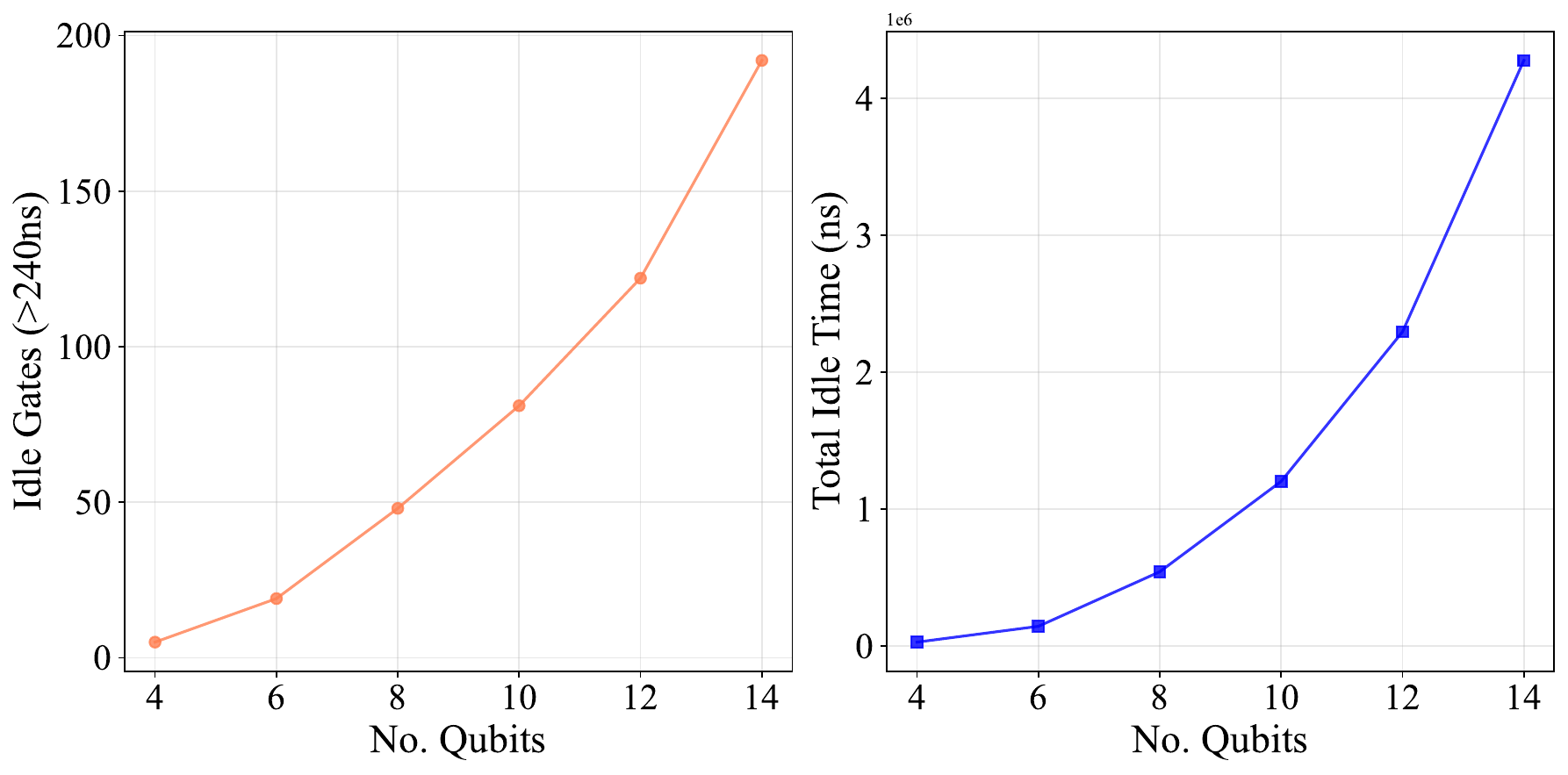}
    \caption{ 
    Analysis of the idle qubits statistics in transpiled QFT circuits as a function of number of qubits. (Left) The total number of idle gates exceeding a $240$\,ns duration threshold. (Right) The total accumulated idle time summed across all qubits in the circuit. 
    }
    \label{fig:fig_idle}
\end{figure}

The QFT involves controlled-phase operations between every pair of $N$ qubits, corresponding—without routing overhead—to $N(N-1)/2$ two-qubit interactions and a comparable number of idle intervals. However, executing this algorithm on hardware with sparse connectivity, such as IBM heavy-hexagonal topology, introduces substantial routing overhead. Since many logical interactions involve non-adjacent qubits, the transpiler must insert SWAP operations to move quantum states across the processor. These additional SWAP operations significantly increasing both the two-qubit gate count and circuit depth. This overhead fragments the schedule, increasing both the number and duration of idle intervals.

Figure~\ref{fig:fig_idle} shows the scaling of the idle qubits statistics in transpiled QFT circuits. As the number of qubits increases, both the total idle time and the number of idle gates scale rapidly.

\begin{table}[h]
    \centering
    \begin{tabular}{c c c c c}
        \toprule
        & \multicolumn{2}{c}{\textit{ibm\_yonsei}} & \multicolumn{2}{c}{\textit{ibm\_fez}} \\
        \cmidrule(lr){2-3} \cmidrule(lr){4-5}
                                 & {35 qubit} & {56 qubit} & {35 qubit} & {55 qubit} \\
        \midrule
        Number of Idle Gates    & 147       & 255       & 48    & 76    \\
        Total Idle Time (ns)    & $1.1 \times 10^6$ & $3.3 \times 10^6$   & $1.7 \times 10^4$ & $3.4 \times 10^4$ \\ 
        \bottomrule
    \end{tabular}
    \caption{
    Idle qubits statistics for the SQD experiments on the \textit{ibm\_yonsei} and \textit{ibm\_fez}. For each system size, the number of idle gates exceeding the thresholds ($240$\,ns for \textit{ibm\_yonsei}, $96$\,ns for \textit{ibm\_fez}) and the corresponding total idle time. 
    }
    \label{tab:sqd_idle}
\end{table}

The LUCJ ansatz used in the SQD experiments exhibits a circuit depth that scales approximately linearly with the number of spatial orbitals. The orbital-rotation blocks contribute dominantly to this scaling, while the Jastrow terms add only a nearly constant depth per layer when sparsity is enforced. Consequently, as the number of orbitals increases, both the number of idle gates and the total idle time grow proportionally.

Table~\ref{tab:sqd_idle} summarizes the idle qubits statistics extracted from the transpiled SQD circuits. On the \textit{ibm\_yonsei}, the number of idle gates increases from $147$ to $255$ as the circuit scales from $35$ to $56$ qubits, with the total idle time increasing from $1.1\times10^6$,ns to $3.3\times10^6$,ns. In contrast, the \textit{ibm\_fez} exhibits substantially fewer idle gates—only 48 to 76—and total idle times shorter by nearly two orders of magnitude, ranging from $1.7\times10^4$ to $3.4\times10^4$,ns. This substantial reduction in idle time implies that \textit{ibm\_fez} experiences significantly less decoherence, consistent with the higher raw-sample fidelities observed in the SQD experiments.

\subsection{Device properties}

All experiments were conducted on IBM superconducting transmon-based quantum processors. The \textit{ibm\_yonsei} was used for both the QFT and SQD experiments presented in the main text, whereas the \textit{ibm\_fez} was employed for additional SQD experiments. Both devices feature a heavy-hexagonal lattice topology, where each qubit is coupled to two or three nearest neighbors to reduce crosstalk while ensuring scalable connectivity.

The \textit{ibm\_yonsei} processor, based on IBM Eagle r3 architecture, comprising $127$ fixed-frequency superconducting transmon qubits coupled via fixed capacitive links. Its basis gate set comprises single-qubit (SQ) rotations $\{R_Z(\theta),\, SX,\, X\}$ and a two-qubit (TQ) echoed cross-resonance (ECR) gate~\cite{sheldon2016procedure}. In combination with appropriate SQ rotations, the ECR gate realizes a CNOT gate and serves as the principal entangling operation on the Eagle devices. SQ gates are implemented with $60$ns pulses shaped using the Derivative Removal by Adiabatic Gaussian (DRAG) pulses~\cite{motzoi2009simple}, while TQ gates are driven by $636$ns ECR pulses. The readout length is $840$ns.

\begin{table*}[h]
    \centering 
    \begin{tabular}{c cccc cccc cccc}
        \toprule
        \multirow{2}{*}{\# Qubits} &
        \multicolumn{4}{c }{$T_1$ ($\mu$s)} &
        \multicolumn{4}{c }{$T_2$ ($\mu$s)} &
        \multicolumn{4}{c }{SQ Gate Error (\%)} \\
        \cmidrule(lr){2-5}\cmidrule(lr){6-9}\cmidrule(lr){10-13}
         & Min & Mean & Median & Max 
         & Min & Mean & Median & Max
         & Min & Mean & Median & Max \\
        \midrule
        4  & 168.75 & 231.34$\pm$61.78 & 221.01 & 314.61 & 
            15.29 & 154.98$\pm$100.52 & 182.02 & 240.58 &
            0.0154 & 0.0170$\pm$0.0019 & 0.0166 & 0.0196 \\
        6  & 168.75 & 236.11$\pm$58.70 & 221.01 & 314.61 &
            15.29 & 173.50$\pm$86.91 & 191.74 & 251.38 &
            0.0154 & 0.0188$\pm$0.0035 & 0.0184 & 0.0249 \\
        8  & 168.75 & 248.07$\pm$57.15 & 242.80 & 317.11 &
            15.29 & 187.36$\pm$85.41 & 191.74 & 294.84 &
            0.0154 & 0.0192$\pm$0.0034 & 0.0184 & 0.0249 \\
        10 & 113.02 & 293.35$\pm$93.31 & 300.92 & 459.76 &
            101.82 & 223.28$\pm$80.87 & 231.91 & 320.79 &
            0.0085 & 0.0297$\pm$0.0237 & 0.0237 & 0.0928 \\
        12 & 110.46 & 268.08$\pm$106.54 & 291.32 & 459.76 &
            105.51 & 223.55$\pm$77.01 & 231.91 & 320.79 &
            0.0085 & 0.0239$\pm$0.0091 & 0.0237 & 0.0368 \\
        14 & 110.46 & 287.07$\pm$104.60 & 316.40 & 459.76 &
            101.82 & 194.28$\pm$80.77 & 176.38 & 320.79 &
            0.0085 & 0.0279$\pm$0.0206 & 0.0233 & 0.0928 \\
        \bottomrule
    \end{tabular}

    \vspace{1.5em} 

    \centering 
    \begin{tabular}{cccc cccc c}
        \toprule
        \multicolumn{4}{c }{TQ Gate Error (\%)} &
        \multicolumn{4}{c }{Readout Error (\%)} &
        \multirow{2}{*}{Physical Qubits Used} \\
        \cmidrule(lr){1-4}\cmidrule(lr){5-8}
         Min & Mean & Median & Max 
         & Min & Mean & Median & Max &  \\
        \midrule
        0.4172 & 0.6290$\pm$0.4114 & 0.4264 & 1.2460 &
             0.1953 & 1.1536$\pm$1.0622 & 0.8789 & 2.6611 &
             {[33, 38, 39, 40]} \\
        0.4172 & 0.6992$\pm$0.4236 & 0.4297 & 1.2460 &
             0.1953 & 1.1637$\pm$0.8817 & 0.8789 & 2.6611 &
             {[33, 37, 38, 39, 40, 41]} \\
        0.4172 & 0.7930$\pm$0.5254 & 0.4330 & 1.7162 &
             0.1953 & 1.9958$\pm$2.6897 & 0.8789 & 8.3740 &
             {[33, 37, 38, 39, 40, 41, 42, 53]} \\
        0.3745 & 0.5906$\pm$0.1784 & 0.5519 & 0.9046 &
             0.3418 & 3.5815$\pm$3.0398 & 2.9419 & 9.6436 &
             {[5, 6, 7, 8, 9, 16, 24, 25, 26, 27]} \\
        0.3745 & 1.0049$\pm$1.0937 & 0.5790 & 4.3089 &
             0.3418 & 2.9236$\pm$2.6729 & 1.9653 & 7.8857 &
             {[4, 5, 6, 7, 8, 9, 10, 16, 25, 26, 27, 28]} \\
        0.3745 & 0.7013$\pm$0.3144 & 0.5938 & 1.6149 &
             0.3418 & 3.0622$\pm$3.0978 & 1.9653 & 9.6436 &
             {[6, 7, 8, 9, 10, 16, 24, 25, 26, 27, 28, 29, 35, 47]} \\
        \bottomrule
    \end{tabular}
    \caption{ 
    Summary of device properties on \textit{ibm\_yonsei} processor used in the QFT experiments.
    }
    \label{tab:qft_properties} 
\end{table*}

Table~\ref{tab:qft_properties} summarizes the device calibration data for all qubits used in the QFT experiments. For each configuration, the table lists the minimum, mean~$\pm$~standard deviation, median, and maximum values of the energy relaxation time ($T_1$), dephasing time ($T_2$), single- and two-qubit gate errors, and readout errors, along with the physical qubits used in each circuit layout. 

\begin{table*}[h]
    \centering
    \begin{tabular}{lc cccc cccc}
        \toprule
        \multirow{2}{*}{Device} & \multirow{2}{*}{\# Qubits} &
        \multicolumn{4}{c }{$T_1$ ($\mu$s)} &
        \multicolumn{4}{c }{$T_2$ ($\mu$s)} \\
        \cmidrule(lr){3-6}\cmidrule(lr){7-10}
         &  & Min & Mean & Median & Max &
              Min & Mean & Median & Max \\
        \midrule
        \multirow{2}{*}{\textit{ibm\_yonsei} (Eagle)} 
         & 35 & 89.80 & 249.29$\pm$87.75 & 230.49 & 429.07 &
                 9.66 & 169.57$\pm$99.69 & 172.82 & 339.47 \\
         & 56 & 34.32 & 245.11$\pm$84.51 & 241.43 & 429.07 &
                 16.54 & 167.58$\pm$97.33 & 169.28 & 392.83 \\
        \midrule
        \multirow{2}{*}{\textit{ibm\_fez} (Heron)} 
         & 35 & 82.25 & 139.07$\pm$32.64 & 134.41 & 196.34 &
                 10.71 & 100.80$\pm$49.23 & 95.41 & 186.61 \\
         & 55 & 53.03 & 141.83$\pm$39.68 & 140.15 & 251.55 &
                 13.02 & 96.95$\pm$55.11 & 92.81 & 239.33 \\
        \bottomrule
    \end{tabular}
    
    \vspace{1.5em} 

    \centering
    
    \begin{tabular}{cccc cccc cccc}
        \toprule
        \multicolumn{4}{c }{SQ Gate Error (\%)} &
        \multicolumn{4}{c }{TQ Gate Error (\%)} &
        \multicolumn{4}{c }{Readout Error (\%)} \\
        \cmidrule(lr){1-4}\cmidrule(lr){5-8}\cmidrule(lr){9-12}
        Min & Mean & Median & Max &
              Min & Mean & Median & Max &
              Min & Mean & Median & Max \\
        \midrule
        0.0156 & 0.0459$\pm$0.0381 & 0.0302 & 0.1848 &
                 0.4239 & 1.1212$\pm$0.7778 & 0.8424 & 3.7370 &
                 0.1709 & 3.6579$\pm$3.4700 & 2.051 & 11.6455 \\
        0.0147 & 0.0453$\pm$0.0360 & 0.0315 & 0.1848 &
                 0.4239 & 1.1167$\pm$0.6825 & 0.8983 & 3.7370 &
                 0.1709 & 3.2048$\pm$3.0234 & 2.002 & 11.6455 \\
        \midrule
        0.0165 & 0.0305$\pm$0.0125 & 0.0269 & 0.0891 &
                 0.1785 & 0.3627$\pm$0.1959 & 0.3027 & 0.9086 &
                 0.3174 & 1.1886$\pm$1.0046 & 0.8545 & 5.9326 \\
        0.0149 & 0.0323$\pm$0.0121 & 0.0295 & 0.0758 &
                 0.1903 & 0.8773$\pm$2.2930 & 0.3104 & 12.0996 &
                 0.2686 & 2.1871$\pm$2.8401 & 1.0254 & 12.1582 \\
        \bottomrule
    \end{tabular}
    \caption{ 
    Summary of device properties on \textit{ibm\_yonsei} and \textit{ibm\_fez} processor used in the SQD experiments.
    }
    \label{tab:sqd_properties} 
\end{table*}

The additional SQD experiments were conducted on the \textit{ibm\_fez} processor, which is based on IBM Heron r2 architecture. The \textit{ibm\_fez} comprises $156$ fixed-frequency transmon qubits connected via \textit{tunable couplers}. Unlike the Eagle architecture, Heron employs the controlled-Z (CZ) gate as its basis two-qubit operation. The introduction of tunable couplers represents a significant architectural improvement, effectively mitigating parasitic interactions and suppressing always-on $ZZ$ crosstalk errors that are intrinsic to fixed-coupling architectures. SQ gates are implemented with $24$ns pulses, while TQ gates have a duration of $68$ns. The readout length is $1560$ns.

\begin{figure*}[t!]
    \centering
    \begin{minipage}{0.49\linewidth}
        \centering
        \includegraphics[width=\linewidth]{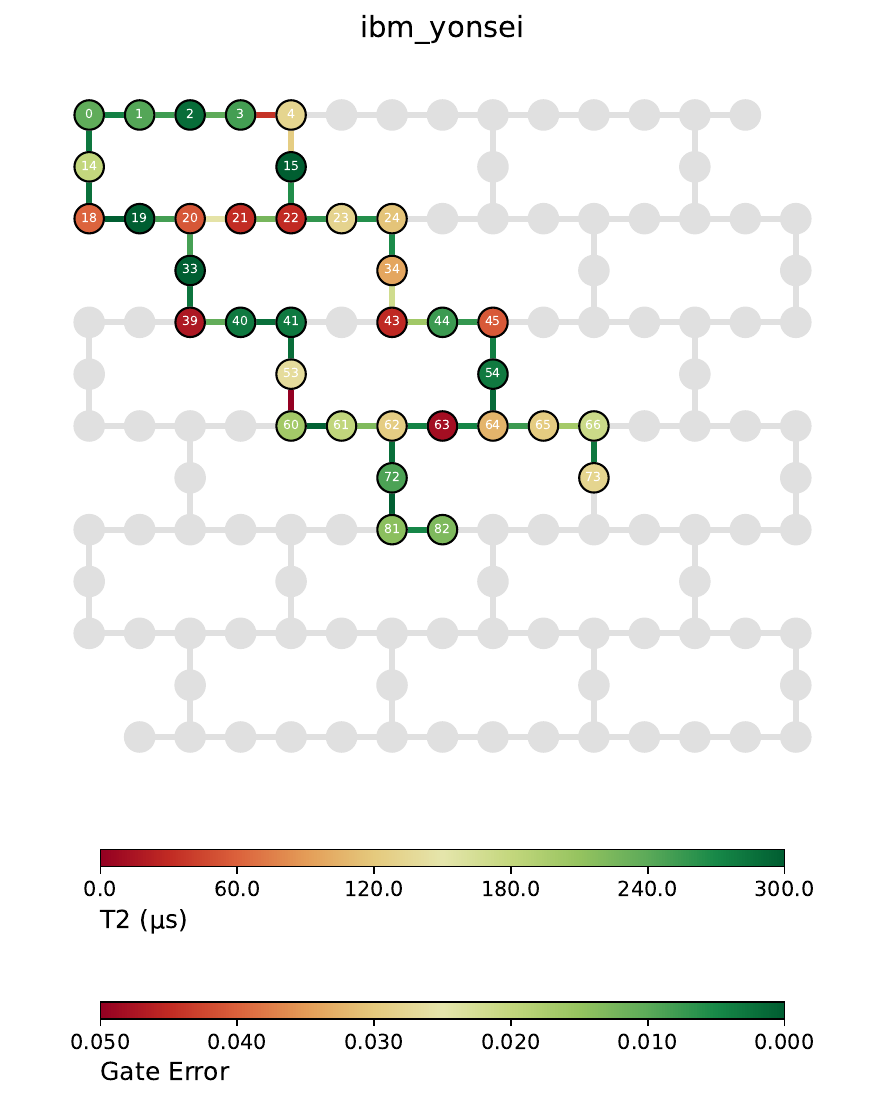}
        (a)
        \label{fig:yonsei_35q}
    \end{minipage}%
    \hfill 
    \begin{minipage}{0.49\linewidth} 
        \centering
        \includegraphics[width=\linewidth]{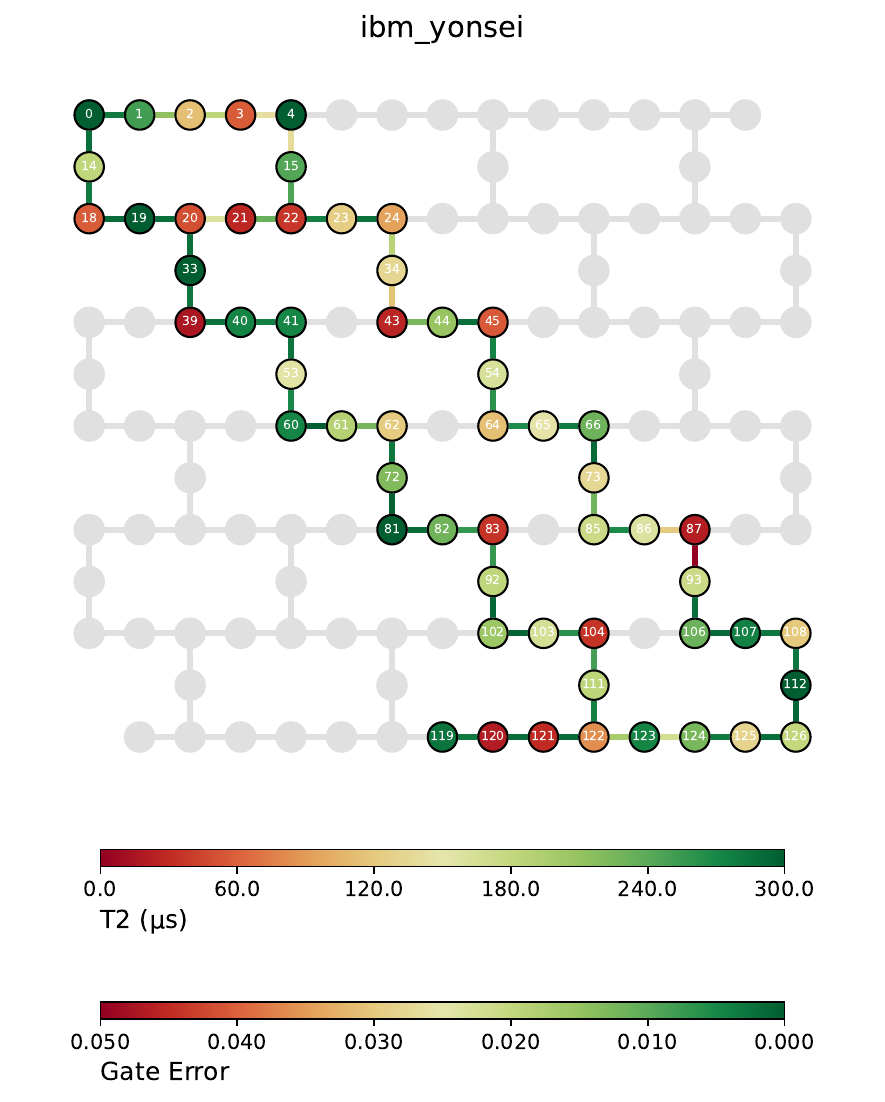}
        (b)
        \label{fig:yonsei_56q}
    \end{minipage}

    \vspace{1em} 

    \begin{minipage}{0.49\linewidth}
        \centering
        \includegraphics[width=\linewidth]{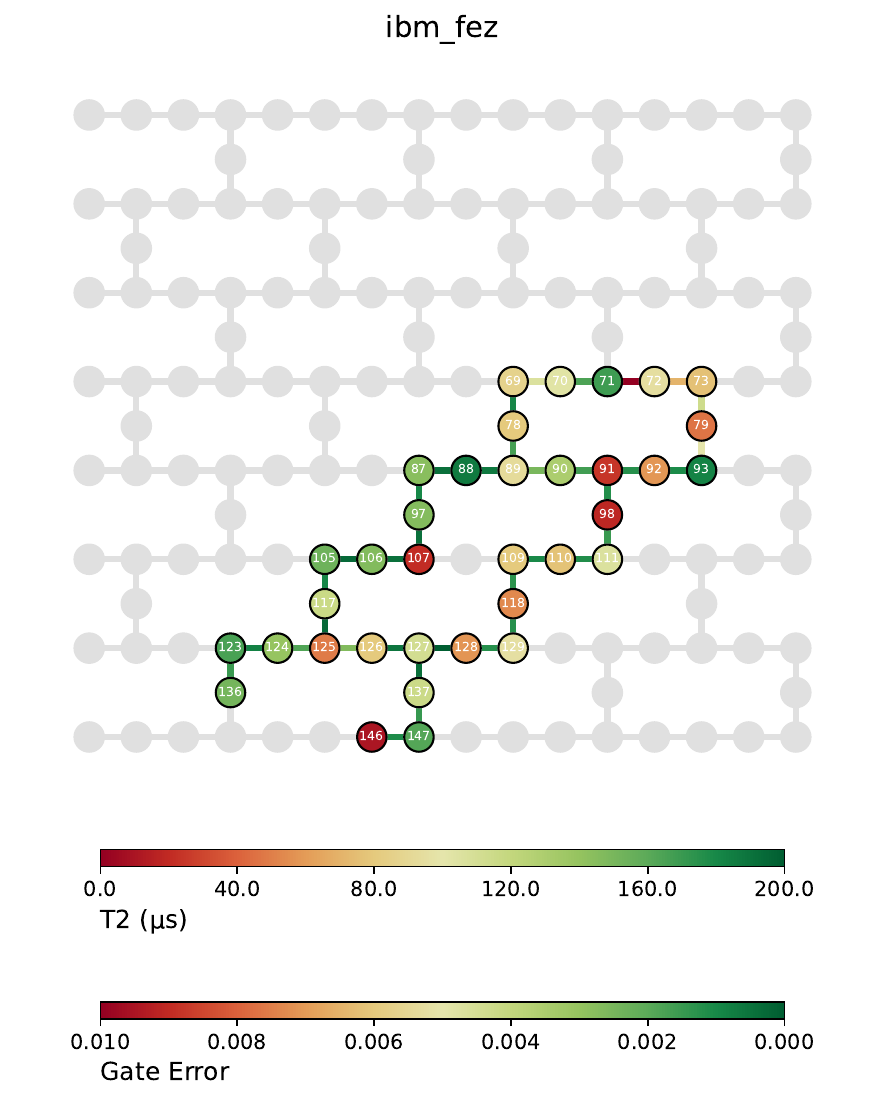}
        (c)
        \label{fig:ibm_fez_35q}
    \end{minipage}%
    \hfill 
    \begin{minipage}{0.49\linewidth} 
        \centering
        \includegraphics[width=\linewidth]{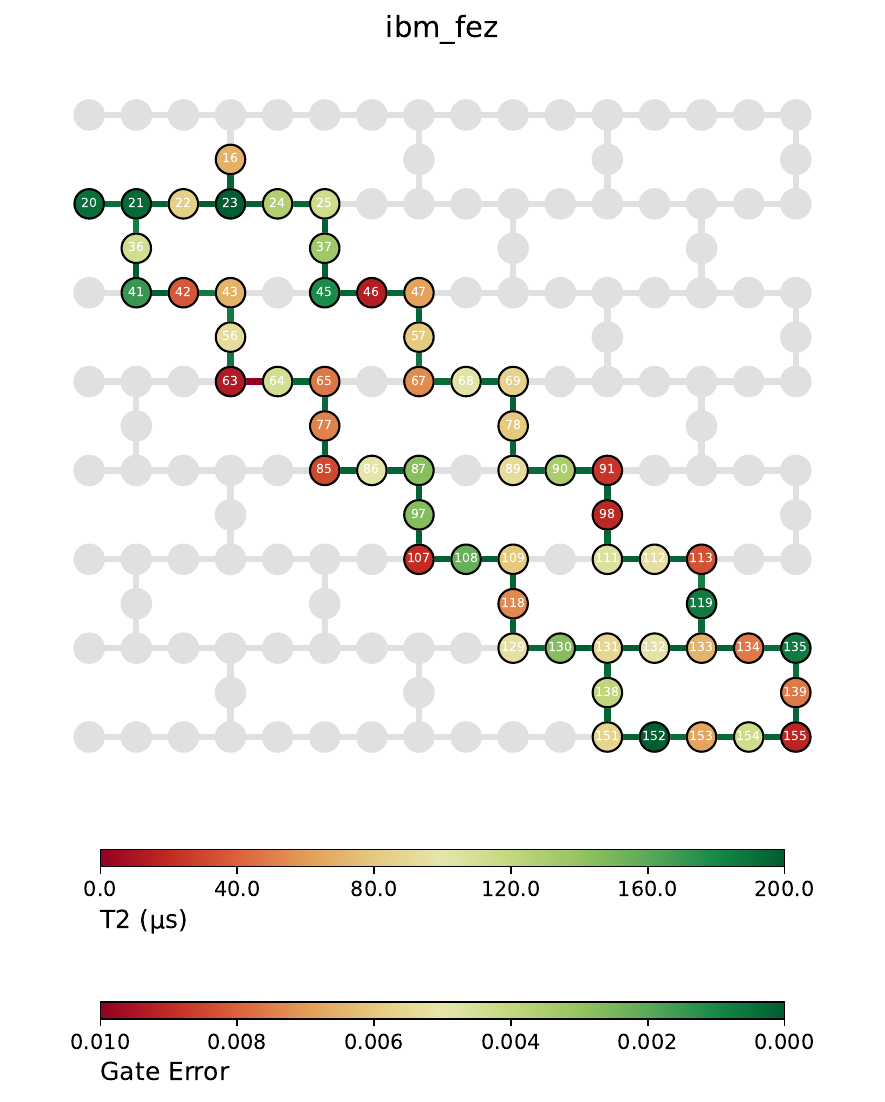}
        (d)
        \label{fig:ibm_fez_55q}
    \end{minipage}

    \caption{
    Physical qubit layouts used for the SQD experiments.
    (a) $35$-qubit layout on the \textit{ibm\_yonsei}.
    (b) $56$-qubit layout on \textit{ibm\_yonsei}.
    (c) $35$-qubit layout on the \textit{ibm\_fez}.
    (d) $55$-qubit layout on \textit{ibm\_fez}.
    }
    \label{fig:all_devices_layout}
\end{figure*}

Table~\ref{tab:sqd_properties} summarizes the device calibration data for all qubits used in the SQD experiments on both \textit{ibm\_yonsei} and \textit{ibm\_fez}. For each configuration, the table lists the minimum, mean~$\pm$~standard deviation, median, and maximum values of the energy relaxation time ($T_1$), dephasing time ($T_2$), single- and two-qubit gate errors, and readout errors. The physical qubits corresponding to each circuit layout are shown in Fig.~\ref{fig:all_devices_layout}.

\end{document}